\documentclass{emulateapj}

\shorttitle{Baryons and dark matter in B1933$+$503}
\shortauthors{Suyu et al.}

\usepackage{natbib}
\usepackage{color}

\newcommand{\boldsymbol}[1]{\mbox{\boldmath{${#1}$}}}

\def\ourlens{B1933$+$503{}}
\def\HST{\textit{HST}}
\def\paramvec{\boldsymbol{\eta}}
\def\param{{\eta}}
\def\data{\boldsymbol{d}}
\def\kms{\rm\,km\,s^{-1}}
\def\kpc{\rm\,kpc}
\def\imfevid{7.2}
\def\mbulge{(1.6\pm0.3)\times10^{10}\,\rm{M_{\sun}}}
\def\RhE{R_{\rm h,E}}
\def\tRhE{\tilde{R}_{\rm h,E}}

\newcommand{\bd}{\begin{displaymath}}
\newcommand{\ed}{\end{displaymath}}
\newcommand{\be}{\begin{equation}}
\newcommand{\ee}{\end{equation}}
\newcommand{\beaa}{\begin{eqnarray*}}
\newcommand{\eeaa}{\end{eqnarray*}}
\newcommand{\bea}{\begin{eqnarray}}
\newcommand{\eea}{\end{eqnarray}}

\begin{document}

\title{Disentangling baryons and dark matter in the spiral
  gravitational lens \ourlens}

\author{S.~H.~Suyu\altaffilmark{1,2,3},
  S.~W.~Hensel\altaffilmark{4,3}, J.~P.~McKean\altaffilmark{5}, 
  C.~D.~Fassnacht\altaffilmark{6}, T.~Treu\altaffilmark{1,*},
  A.~Halkola\altaffilmark{7}, M.~Norbury\altaffilmark{8}, 
  N.~Jackson\altaffilmark{9}, P.~Schneider\altaffilmark{3}, 
  D.~Thompson\altaffilmark{10}, M.~W.~Auger\altaffilmark{1},
  L.~V.~E.~Koopmans\altaffilmark{11},
  K.~Matthews\altaffilmark{12}}
\altaffiltext{1}{Department of Physics, University of California,
  Santa Barbara, CA 93106-9530, USA} 
\email{suyu@physics.ucsb.edu}
\altaffiltext{2}{Kavli Institute for Particle Astrophysics and Cosmology, Stanford University, 452 Lomita Mall, Stanford, CA 94035-4085, USA} 
\altaffiltext{3}{Argelander-Institut f\"ur Astronomie, Auf dem H\"ugel 71, 53121 Bonn, Germany}
\altaffiltext{4}{Max-Planck-Institut f\"ur Mathematik, Vivatsgasse 7,
  53111 Bonn, Germany}
\altaffiltext{5}{ASTRON, Oude Hoogeveensedijk 4, 7991 PD Dwingeloo,
  the Netherlands}
\altaffiltext{6}{Department of Physics, University of California
  Davis, 1 Shields Avenue, Davis, CA 95616, USA}
\altaffiltext{7}{Excellence Cluster Universe, Technische Universit\"at M\"unchen, Boltzmannstr. 2, 85748 Garching, Germany}
\altaffiltext{8}{Las Cumbres Observatory Global Telescope Network, 6740 Cortona Drive, Suite 102, Santa Barbara, CA 93117, USA}
\altaffiltext{9}{Jodrell Bank Centre for Astrophysics, School of
  Physics and Astronomy, University of
  Manchester, Turing Building, Oxford Road, Manchester M13 9PL, UK}
\altaffiltext{10}{Large Binocular Telescope Observatory, University of
  Arizona, 933 North Cherry Avenue, Tucson, AZ 85721, USA}
\altaffiltext{11}{Kapteyn Astronomical Institute, PO Box 800, 9700 AV Groningen, the Netherlands}
\altaffiltext{12}{Caltech Optical Observatories, California Institute of Technology, Pasadena, CA 91125, USA}
\altaffiltext{*}{Packard Research Fellow}

\begin{abstract}

Measuring the relative mass contributions of luminous and dark matter
in spiral galaxies is important for understanding their formation and
evolution.  The combination of a galaxy rotation curve and strong
lensing is a powerful way to break the disk-halo degeneracy that is
inherent in each of the methods individually.  We present an analysis of
the 10-image radio spiral lens \ourlens\ at $z_{\rm l}=0.755$, 
incorporating (1) new global
VLBI observations, (2) new adaptive-optics assisted K-band imaging,
(3) new spectroscopic observations for the lens galaxy rotation curve
and the source redshift.  We construct a three-dimensionally
axisymmetric mass distribution with 3 components: an exponential
profile for the disk, a point mass 
for the bulge, and an NFW profile for the halo.  The mass model is
simultaneously fitted to the kinematics and the lensing data.  The NFW
halo needs to be oblate with a flattening of
$a/c=0.33^{+0.07}_{-0.05}$ to be consistent with the radio data.  This
suggests that baryons are effective at making the halos oblate near
the center.  The lensing and kinematics analysis probe the inner
$\sim$$10\kpc$ of the galaxy, and we obtain a lower limit on the halo
scale radius of $16\kpc$ (95\% CI).
The dark matter mass fraction inside a sphere with a radius of 2.2
disk scale lengths is $f_{\rm DM,2.2}=0.43^{+0.10}_{-0.09}$.
The contribution of the disk to the total circular velocity at 2.2
disk scale lengths is $0.76^{+0.05}_{-0.06}$, suggesting that the disk
is marginally submaximal.  The stellar mass of the disk from our modeling is
$\log_{10}(M_{*}/{\rm M}_{\sun}) = 11.06^{+0.09}_{-0.11}$
assuming that the cold gas contributes $\sim$$20\%$ to the total disk
mass.  In comparison to the stellar masses estimated from stellar
population synthesis models, the stellar initial mass function of
Chabrier is preferred to that of Salpeter by a probability factor of 7.2.

\end{abstract}

\keywords{galaxies: halos --- galaxies: individual (\ourlens) ---
  galaxies: kinematics and dynamics --- galaxies: spiral ---
  gravitational lensing } 

\section{Introduction}

The discovery of flat rotation curves near and beyond the optical edge
of galaxies provides strong evidence for the existence of dark matter
\citep{Rubin++78, Bosma78, vanAlbadaSancisi86}.  Since then, observations
of the cosmic 
microwave background, supernovae, galaxy clusters, weak lensing,
baryon acoustic oscillations, and gravitational lens time delays indicate
that our Universe is well 
described by a model comprised of cold dark matter and dark energy
\citep[see, e.g.,][]{Komatsu++11, Suzuki++11, Mantz++10,
  Schrabback++10, Blake++11, Suyu++10}.  Even though the $\Lambda$-CDM model is
successful at explaining the Universe on large scales, the interplay
between dark matter and baryons on galactic scales remains an open
question.  

$N$-body simulations of dark matter particles show that equilibrium
dark matter halos have spherically averaged mass density profiles that
are nearly universal \citep[NFW;][]{Navarro++96} and are typically triaxial in
shape \citep[e.g.,][]{JingSuto02}.  The inclusion of baryons in
simulations is challenging due to the large dynamical range in mass
and uncertainties in the cooling and feedback mechanisms.  Using a
subset of the OverWhelmingly Large Simulations project
\citep{Schaye++10} that included various prescriptions of cooling and
feedback, \citet{Duffy++10} found that the inner profile of
galaxy-scale dark matter halos is very sensitive to the baryon
physics.  With weak stellar feedback from supernovae, the inner
profiles tend to steepen and become more isothermal as a result of the
high central baryon fractions that pulls the dark matter towards the
center.  This ``halo contraction'' is also found in other studies
\citep[e.g.,][]{Blumenthal++86, Gnedin++04, Gnedin++11}.  On the other
hand, with strong feedback  
from both supernovae and an
active galactic nucleus, the inner profile is very similar to
the NFW profile from dark-matter only simulations.  Measuring the
inner profiles of dark matter halos therefore helps determine the
kinds of baryonic processes that occur during galaxy formation and
evolution.

Observationally, probing the inner profiles of dark matter halos with
rotation curve data is difficult due to the disk-halo degeneracy
\citep[e.g.,][]{vanAlbadaSancisi86, Dutton++05}. Since the rotation curve 
primarily depends on the \textit{total} enclosed mass within spherical
radii, a heavy disk with a light halo and a light disk with a heavy
halo can both be fit to the rotation curve.  The degeneracy is 
prominent in models where the disk and halos have fixed parametric
forms, and is reduced
in self-consistent models where the halo shape changes in
response to the presence of the disk \citep{AmoriscoBertin10}.
To circumvent the disk-halo degeneracy, 
some studies have assumed that the disk contributes maximally to the
circular velocity.  However, studies based on the Tully-Fisher
relation or the central vertical velocity dispersion of disk stars
have shown that disks tend to be submaximal
\citep[e.g.,][]{CourteauRix99, Bottema93, Bershady++11}.  To break the 
disk-halo degeneracy without resorting to maximal-disk assumptions,
one needs to measure independently the relative mass contribution of
the disk and the dark matter halo, or equivalently, the mass-to-light
ratio ($M/L$) of the disk.  Stellar population synthesis (SPS) models
allows estimations of the stellar mass and hence the $M/L$ of the disk.
However, uncertainties in the stellar mass due to, for example, the
unknown stellar initial mass function (IMF), limits the accuracy of
this approach \citep[e.g.,][]{Conroy++09}.  Therefore, disentangling
the contributions of the disk and the halo to the rotation curve is
key in understanding both the inner halo profile and the stellar IMF.

An effective way to overcome the disk-halo degeneracy is to combine
rotation curves with strong gravitational lensing.  If a spiral galaxy lies
along the line of sight between the observer and a background source,
the source can be strongly lensed into multiple images by the spiral
galaxy \citep[e.g.,][]{Treu10}.  While kinematics probe mass within
spheres, strong lensing 
probes mass enclosed within cylinders (within the ``Einstein radius'',
which is roughly the radial distance of the multiple images from the
lens galaxy center).  The combination of the two methods with different
mass dependence breaks the disk-halo degeneracy.  
Spectroscopic and
imaging surveys in recent years have substantially enlarged the
sample of spiral lenses, totaling more than $20$ now
\citep[e.g.,][]{Feron++09, Sygnet++10, Treu++11}.  The first few analyses of
spiral lenses are already informing us about disk-maximality and the
stellar IMFs in these systems \citep[e.g.,][]{Koopmans++98,
  Maller++00, Trott++10, Dutton++11}.

In this paper, we study \ourlens, a spiral gravitational lens at
$z_{\rm l}=0.755$ with 10
radio lensed images that was discovered by \citet{Sykes++98} in the
Cosmic Lens All-Sky Survey \citep[CLASS;][]{Myers++03, Browne++03}. 
Previous modeling of the radio data by \citet{Cohn++01} tested
power-law models for the combined dark and baryonic mass distribution
of the lens.  Here,  
we obtain new radio, infrared and spectroscopic observations, and
construct a 3-component mass model (for the disk, bulge and dark
matter halo) that is simultaneously fitted to
both the kinematics and lensing data.  The analysis is very similar in
spirit to the one presented by \citet{Dutton++11} on the lens
SDSS\,J2141$-$0001, except we use an NFW instead of an isothermal
profile to describe the dark matter halo.  The aims of our study are
(1) measure the inner shape and profile of the dark matter halo, (2)
determine the relative contributions of the disk, bulge and halo, and
(3) place constraints on the stellar IMF.

The paper is organized as follows.  In Section \ref{sec:obs}, we
present observations of the radio lens \ourlens.  The alignment of the
radio and near-infrared images is described in Section
\ref{sec:align}, and the lens light profile in the near-infrared image
is measured in 
Section \ref{sec:lenslight}.  The 3-component mass model is outlined
in Section \ref{sec:mass-model}, and the statistical framework for the
analysis is described in Section \ref{sec:Bayes}.  We present the
kinematics-only and lensing-only results in Sections \ref{sec:kinematics}
and \ref{sec:lensing}, respectively.  We discuss the results and
implications of the joint kinematics and lensing analysis in Section
\ref{sec:L+D}, before concluding in Section \ref{sec:conclude}.

Throughout the paper, we assume a flat $\Lambda$-CDM cosmology with
$H_0=70\,{\rm km\,s^{-1}\,Mpc^{-1}}$, $\Omega_{\Lambda}=1-\Omega_{\rm m}=0.73$.
In this cosmology, $1''$ corresponds to $7.5\kpc$ at the lens redshift
and $8.7\kpc$ at the source redshift, which is measured in Section
\ref{sec:obs:spec:z}.  Images of the lens system are 
north up and east left.  Parameter constraints are given as the median
values with uncertainties given by the 16th and 84th percentiles
(corresponding to 68\% credible intervals (CI)) unless otherwise stated.  

\section{Observations}
\label{sec:obs}

We obtained both lensing and kinematic observations of \ourlens\ for
constraining the lens mass distribution.  We present the global VLBI
observations of the lensed radio source in Section
\ref{sec:obs:radio}, near-infrared imaging of the lens system in
Sections \ref{sec:obs:AO} and \ref{sec:obs:nirc}, and spectroscopic data
sets for obtaining the rotation 
curve of the lens galaxy and the source redshift in Sections
\ref{sec:obs:spec:ESI} and \ref{sec:obs:spec:z}.  We describe the
archival \textit{Hubble Space 
  Telescope} (\HST) images in Section \ref{sec:obs:HST}.

\subsection{Radio observations}
\label{sec:obs:radio}

We observed \ourlens\ with the global Very Long Baseline
Interferometry (VLBI) array on 1998 November 27 
at 1.7\,GHz with a bandwidth of 16\,MHz.  We used 17 telescopes
with 10 from the Very Long Baseline Array 
(VLBA) and 7 from the European VLBI Network (EVN). We adopt the center
of component 4 of the lensed images as the phase center for the
observations.  Observations were conducted on a cycle of 6.5 minutes,
with 1.5 minutes on a phase calibrator (B1954$+$51=J1955$+$5131) followed
by 5 minutes on the target source \ourlens. The exception was the
Lovell telescope, which has a slower slew rate and for which every
other observation of the phase calibrator was omitted, yielding a
1.5-minute$+$11.5-minute cycle on the phase reference and target. 
The total observing time was 9 hours, providing the good $u$-$v$ coverage
required for high dynamic range imaging and image fidelity.  
A single observation of 3C345 was obtained for fringe finding and flux
calibration.

\begin{figure}
  \centering
   \includegraphics[width=0.5\textwidth, clip]{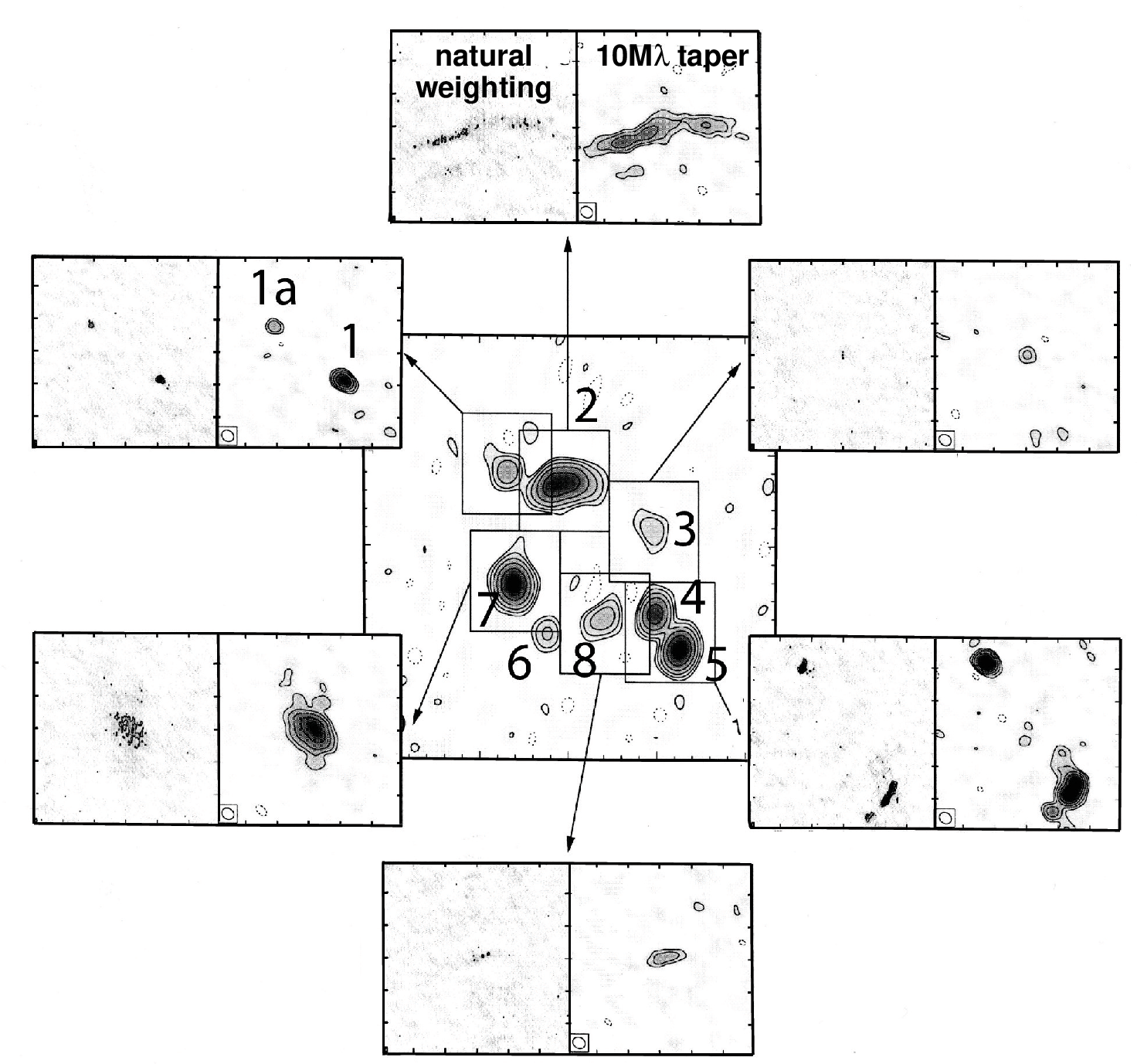}
  \caption{\label{fig:globalVLBI} Global VLBI observations of
    \ourlens.  Center: 1.7 GHz MERLIN observations of \ourlens\ taken
    from \citet{Sykes++98}.  Contours are plotted at $-3$, 3, 6, 12,
    24, 48 times the rms noise level of 150 $\mu$Jy beam$^{-1}$.  The
    beamsize is $139 \times 113$ mas at a position angle of
    $-13.8\degr$.  In each of the insets, the left and right panels
    show the 1.7 GHz global VLBI observations with natural weighting
    and with a 10 M$\lambda$ taper, respectively.  In the left (right)
    panels, contours are plotted at $-3$, 3, 6, 12, 24, 48, 96 times
    the rms noise level of 50 (55) $\mu$Jy beam$^{-1}$, and the
    beamsize is $5.7 \times 2.7$ mas ($20.1\times16.5$ mas) at a
    position angle of $-3.22\degr$ ($60.55\degr$).  Each of the global
    VLBI panels
    covers a 300 mas by 300 mas area with the tick marks separated by
    50 mas.}
\end{figure}

We reduce the data with the Astronomical Image Processing System
(AIPS\footnote{developed by the National Radio Astronomy Observatory})
package using standard procedures. The images are iteratively 
CLEANed and self-calibrated (phase-only), before a single amplitude
self-calibration solution is performed, to remove residual phase and
amplitude errors. The final maps are produced using natural-weighting 
of the visibilities and with a 10 M$\lambda$ taper to increase the
sensitivity to the extended emission. The natural-weighting image has
a root-mean-square (rms) noise level of 50~$\mu$Jy\,beam$^{-1}$ with a
beam size of 5.7~$\times$~2.7\,mas at a position angle (PA) of
$-$3.22$\degr$. The 
tapered image has a rms of 55~$\mu$Jy\,beam$^{-1}$ with a beam size
of 20.1~$\times$~16.5\,mas at a PA of 60.55$\degr$. 
Figure \ref{fig:globalVLBI} shows the final maps for each of the
components in \ourlens, except for component 6 that is not detected.

The radio source being lensed has a compact core with two
extended radio lobes on opposite sides of the core
\citep{Sykes++98, Nair98}.  The radio spectral indices make the
identification of the images of the compact core unambiguous
\citep{Sykes++98}, and the flux densities of the core components
showed little time variability \citep{Biggs++00}.  
Components 1, 3, 4 and 6 correspond to the
images of the compact core, components 1a and 8 are the images of one
of the lobes, and components 2, 5 and 7 correspond to the images of
the other radio lobe.  With component 2 counting as two (merging)
images, \ourlens\ is a spectacular 10-image radio lens.

We use components 1 and 4 to align the global VLBI data to the
previous radio observations (with VLA, MERLIN and VLBA) in
\citet{Sykes++98} and \citet{Marlow++99}.  
Table \ref{tab:image-positions} lists the image
positions for each of the components in the previous observations
compiled by \citet{Cohn++01} and in the global VLBI data.  The merging
pair of images are denoted by components 2a and 2b.  In the global
VLBI data, several components (1, 4, 5 and 2a) have
multiple intensity peaks, but some of their corresponding images (3
and 7) have only single peaks due to, e.g., scatter-broadening during
propagation through the disk of the lens galaxy
\citep[e.g.,][]{Marlow++99, Norbury02}.  We adopt the 
flux weighted average of the peak positions as the image positions and
estimate the uncertainties from the separation of the peaks.  We also
set the positional uncertainty of component 7 and 2b to that of their
counter image (component
5).  For components that are spatially extended (1a and 8), the
uncertainty is set to the geometric mean of the major and minor axis
of the beam.

\citet{Cohn++01} also compiled flux ratios from \citet{Sykes++98},
\citet{Nair98}, and \citet{Biggs++00}.  We do not list or use the flux
ratios because they could be systematically biased due to scattering
and substructures in the lens.  In fact,
  \citet{KochanekDalal04} showed that the flux ratios in \ourlens\ are
  anomalous due to substructures, and the constraints in \ourlens\ lead
  to a relatively smooth, elliptical macro mass model.
Furthermore, \citet{Cohn++01} found that the exclusion of flux ratios
had little effect on the main results of their mass models.

\begin{table*}
\begin{center}
\caption{Radio Image Positions}
\label{tab:image-positions}
\begin{tabular}{ccccccccccc}
\hline
 & & \multicolumn{4}{c}{\citet{Cohn++01}} & & \multicolumn{4}{c}{global VLBI} \\
Component & & System & $\Delta$R.A. & $\Delta$Dec & Uncertainty & & System & $\Delta$ R.A. & $\Delta$ Dec & Uncertainty \\
ID & & ID & (arcsec) & (arcsec) & (arcsec) &  & ID & (arcsec) & (arcsec) & (arcsec) \\ \hline
1a & & I  & $\phantom{-}0.545$ & $\phantom{-}0.584$ & $0.02\phantom{0}$  & & V  & $\phantom{-}0.5538$ & $\phantom{-}0.5774$ & 0.0042\\
8  & &    &  $-0.114$          & $-0.335$           & $0.02\phantom{0}$  & &    & $-0.1219$           & $-0.3266$           & 0.0042\\ \hline
1  & & II & $\phantom{-}0.447$ & $\phantom{-}0.495$ & 0.001 & & VI & $\phantom{-}0.4459$ & $\phantom{-}0.4945$ & $0.001\phantom{0}$\\
3  & &    & $-0.389$           & $\phantom{-}0.158$ & 0.001 & &    & $-0.3874$           & $\phantom{-}0.1635$ & 0.0023\\
4  & &    & $-0.397$           & $-0.299$           & 0.001 & &    & $-0.3959$           & $-0.2985$           & $0.001\phantom{0}$\\
6  & &    & $\phantom{-}0.230$ & $-0.387$           & 0.005 & &    & --- & --- & \phantom{--}---\\ \hline
5  & & III& $-0.531$           & $-0.497$           & 0.005 & &    & --- & --- & \phantom{--}---\\
7  & &    & $\phantom{-}0.398$ & $-0.134$           & 0.005 & &    & --- & --- & \phantom{--}---\\ \hline
2a & & IV & $\phantom{-}0.189$ & $\phantom{-}0.412$ & 0.072 & & VII& $\phantom{-}0.1894$ & $\phantom{-}0.4129$ & $0.04\phantom{00}$\\
2b & &    & $\phantom{-}0.061$ & $\phantom{-}0.425$ & 0.042 & &    & $\phantom{-}0.0737$ & $\phantom{-}0.4296$ & $0.007\phantom{0}$\\
5  & &    & $-0.522$           & $-0.514$           & 0.045 & &    & $-0.5310$           & $-0.4922$           & $0.007\phantom{0}$\\
7  & &    & $\phantom{-}0.417$ & $-0.130$           & 0.049 & &    & $\phantom{-}0.3970$ & $-0.1308$           & $0.007\phantom{0}$\\
\hline
\end{tabular} 
\end{center}
Notes.  Column 1 lists the radio components as shown in Figure
\ref{fig:globalVLBI}.  Components 5 and 7 appear twice, with the peaks
forming a two-image system, and the extended features forming a
four-image system with the merging components 2a and 2b.  Columns 2--5
are from \citet{Cohn++01}, and columns 6--9 are from the global VLBI
observations in this paper.  Columns 2 and 6 are the ID numbers for
each multiple image system.  Columns 3 and 7 (4 and 8) are the relative
right ascensions (declinations) in the coordinate system of
\citet{Cohn++01} where the origin is close to the lens center.  
Columns 5 and 9 are the estimated positional 
uncertainties.
\vspace{0.2cm}
\end{table*}

\subsection{Keck AO-assisted NIRC2 image}
\label{sec:obs:AO}

We observed \ourlens\ with the adaptive optics (AO) system at the Keck II
telescope on 2005 July 31.  We used the Kp filter (at 2.2\,$\mu$m) on the Near
Infrared Camera 2 (NIRC2), 
and took 33 dithered frames with a field of view of $10''\times10''$
that are each $6$ co-adds of $30$ 
seconds.  The dithering allows good sky subtraction in the reduction.
The natural tip-tilt reference star of magnitude $R$$\sim$$16.4$
located at $\sim$$21\farcs3$ southeast of \ourlens\ and a sodium laser guide
star were used to correct for atmospheric turbulence. 

\begin{figure}[t]
  \centering
  \includegraphics[width=0.4\textwidth]{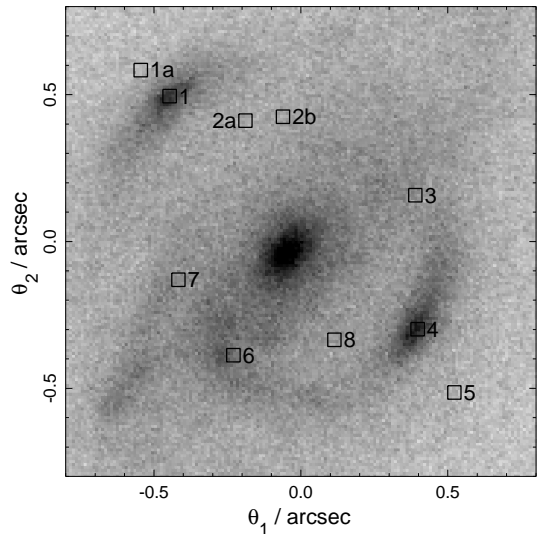}
  \caption{\label{fig:AOimage} NIRC2 Kp image of \ourlens.  
   The radio image positions from \citet{Cohn++01} are overlaid.  The
   alignment of the NIRC2 and radio images are described in Section
   \ref{sec:align}.  The
   lensing arcs are associated with radio components 1, 3, 4 and 6, which
   are images of the compact core in the source.  The arcs
   corresponding to components 3 and 6 are faint due to severe dust
   extinction in the plane of the spiral galaxy.}
\end{figure}

We use an IRAF-based algorithm to subtract the sky, remove bad pixels
and cosmic rays, and produce weight maps for each frame.  We use the
weight maps and the {\sc Drizzle} package \citep{FruchterHook02} to coadd
the images together.  We show in Figure \ref{fig:AOimage} the drizzled
NIRC2 image of the lens system.  By comparing the positions of the three
stars in the NIRC2 image to the corresponding stars in previous
\textit{Hubble Space Telescope} (\HST) observations of the system
(Section \ref{sec:obs:HST}), we determine the pixel scale of the NIRC2
image to be $9.964$\,mas\,pix$^{-1}$.  The closest field star
to \ourlens\ ($\sim$$2\farcs9$ away) in the combined NIRC2 
image has a full width at half-maximum (FWHM) of $\sim$57\,mas.

\subsection{Keck NIRC image}
\label{sec:obs:nirc}

In addition to the high-resolution NIRC2 image, we observed \ourlens\
on 1996 July 31 with the Near Infrared Camera
\citep[NIRC;][]{MatthewsSoifer94} at the Keck Observatory.  The NIRC
detector is 256 pixels on a side, with a pixel scale of 0\farcs15
pix$^{-1}$.  Thus, the field of view of the camera is 38\farcs4 on a
side, i.e., larger than the NIRC2 narrow camera field of view.  The
system was observed in both the J and K bands, but the image quality
of the J-band data was too poor to be useful.  The K-band data consist
of 59 exposures, each with an exposure time of one minute (5 co-adds of
12 s each).  The data were processed in a standard fashion, including
steps to subtract the dark current, flatten the images, and subtract
the sky.  For each of the images, the flat-field and sky frames were
constructed from the frames observed directly before and after the
image.  The processed images were aligned by measuring the position of
a star that appeared in each frame and then were co-added to produce
the final image.  The photometric zero point of the NIRC data was
determined through a comparison to the Two
Micron All Sky Survey \citep[2MASS;][]{Skrutskie++06}, which has
K-band magnitudes for the two brightest stars in the image.  The
uncertainty in the photometric zero point is estimated to be 0.15 by
comparing the magnitudes of the two stars.  This incorporates the
uncertainty due to the difference in throughput of the 2MASS and NIRC
K-band filters, which we estimate to be $\lesssim0.03$ by comparing
the magnitude difference between the filters for several stars.

\subsection{Keck ESI spectroscopy}
\label{sec:obs:spec:ESI}

We observed \ourlens\ using the Echelle Spectrograph and Imager \citep[ESI;][]{SheinisEtal02}
instrument at the Keck Observatory on 2002 June 6.  The slit of width $1\farcs25$ 
was aligned along the major axis of the lens galaxy (PA=$138\degr$).
The standard echellette mode yielded a spectral resolution with
Gaussian width of $30\kms$.  Five exposures of 1800s each were
taken under good seeing conditions of $0\farcs6$.

\begin{figure}
  \centering
  \includegraphics[width=0.35\textwidth,angle=270]{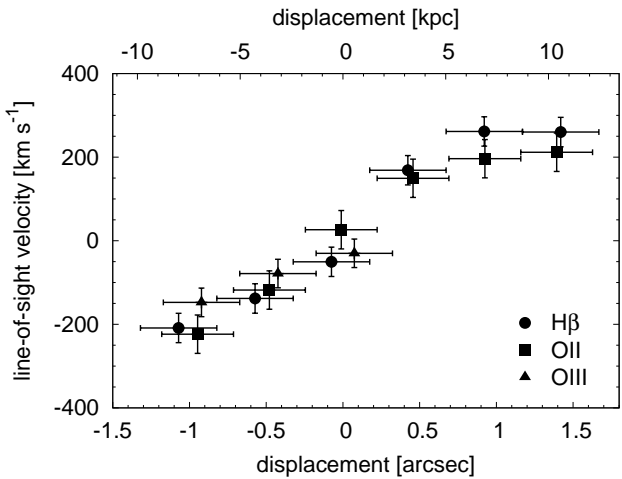}
  \caption{\label{fig:rotcurve} Rotation curve based on the observed
  emissions lines H$\beta$, [O {\sc ii}], and [O {\sc iii}] in the Keck ESI
  observations.  }
\end{figure}

We reduce the data using the software EASI2D developed by D.~J.~Sand
and T.~Treu \citep{Sand++04}.  We see prominent emission lines
H$\beta$ ($\lambda$4861), [O {\sc ii}] ($\lambda$3726, $\lambda$3729), and
[O {\sc iii}] ($\lambda$4959, $\lambda$5007) from the lens galaxy in the
spectra.
For each line, we set the center of the galaxy to the brightest pixel
in the 2-dimensional spectrum since measuring the center to much less
than a pixel is difficult due to the presence of seeing.  We then bin  
the spectrum in the spatial direction by a factor of 3 to increase the
signal. Thus, the center of the galaxy should be well within the
central binned spatial pixel of size $\sim0\farcs5$.  Any small
offsets in the centroid are taken into account in the modeling in
Section \ref{sec:kinematics:results}.
The systemic velocity is difficult to measure from
the data directly and is determined in the modeling
(Section \ref{sec:kinematics:results}).  Figure \ref{fig:rotcurve} is  
the rotation curve based on these lines.

\subsection{Keck NIRSPEC spectroscopy}
\label{sec:obs:spec:z}

We carried out infrared spectroscopic observations of the \ourlens\ 
lensed source on 2005 September 9 with the Near Infrared Spectrometer
\citep[NIRSPEC;][]{McLeanEtal98} on the Keck II telescope. The data were taken through the
NIRSPEC 6 and 7 filters (N6 and N7, respectively), which gave an approximate
wavelength coverage of 1.56 to 2.02~$\mu$m and 2.08 to 2.52~$\mu$m,
respectively. Each observation consisted of 4 $\times$ 300 s
exposures, that used dithering along the slit to improve the removal
of the sky background during the reduction stage. The total exposure
time was 1 hour through each filter. The slit was put at a position
angle of 226.7$\degr$ to cover the two strong infrared components from
the lensed source. The standard star HD162208 was also
  observed four times each in N6 and N7 for calibration.

\begin{figure}[t]
  \centering
  \includegraphics[width=0.53\textwidth]{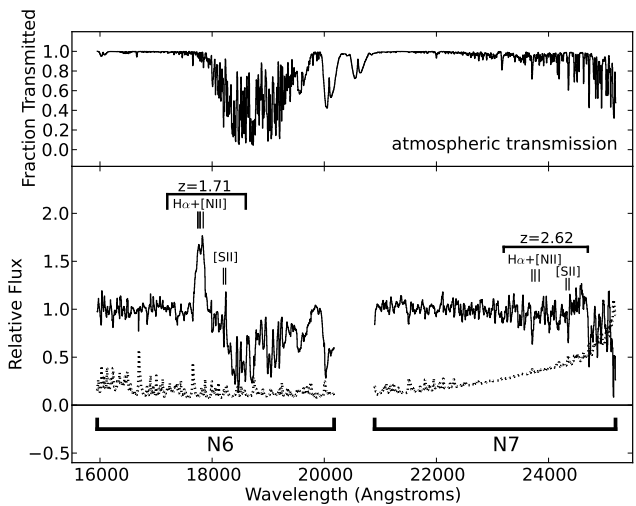}
  \caption{\label{fig:spec:zs} Spectra of the lensed source from
  NIRSPEC on Keck II in the N6 (left) and the N7 (right) filters are
  shown by the solid curves in the bottom panel.  The rms noise
  spectra are the dotted curves.  The spectra are smoothed using a
  7-pixel moving average with each point being weighted by the inverse
  variance associated with it.  The atmospheric transmission is
  shown in the top panel.  In each of the N6 and N7 spectra, five
  emission lines (the first [N {\sc ii}] line, H$\alpha$, the second
  [N {\sc ii}] line and the two [S {\sc ii}] lines) are marked.  The
  set in N6 is for a source redshift of $z_{\rm s}=1.71$, and the
  set in N7 is for $z_{\rm s}=2.62$ (with the broad line in N6
  corresponding to H$\beta$).  The absence of emission lines in N7
  rules out the previously identified $z_{\rm s}=2.62$.}
\end{figure}

We reduce the data with a Python-based pipeline that removes cosmic
rays, subtracts the sky, wavelength calibrates using the atmospheric
sky-lines and extracts a one-dimensional spectrum. Furthermore, we use
the standard star HD162208 to correct for the response of the
spectrograph. The reduced 
spectra for both filters are presented in the bottom panel of Figure
\ref{fig:spec:zs}, and the atmospheric transmission is shown in the
top panel.  The spectra are smoothed using a 7-pixel moving average with
each point being weighted by the inverse variance associated with
it. 
We identify a strong emission line in the N6 spectrum (bottom panel,
left), but do not see any spectral features in the N7 spectrum (bottom
panel, right).  Therefore, we
believe that the detected emission line is likely H$\alpha$ blended
with two [\mbox{N {\sc ii}}] emission lines, corresponding to a source redshift
of $z_{\rm s} = 1.71 \pm 0.01$. The uncertainty in the redshift is
conservative and accounts for the blending of the lines and also the
contamination by atmospheric lines on both sides of the emission line. 

We note that the strong emission line detected here has been
previously reported by \citet{Biggs++00}, based on an unpublished
spectrum that was taken with the United Kingdom Infrared Telescope
(UKIRT). Then the line was interpreted to be H$\beta$ at redshift
2.62, due to a second spectral feature that was
believed to be H$\alpha$ in the K-band. Our much better spectral
resolution and higher sensitivity data do not detect the second
emission line in N7, which would certainly have been detected if it
were H$\alpha$ given the relative flux of the supposed H$\beta$ line in
N6 and the noise level.  We therefore rule out a redshift of 2.62
for the lensed source.

\subsection{Archival \HST\ images}
\label{sec:obs:HST}
\begin{figure}[t]
\centering
\includegraphics[width=0.45\textwidth]{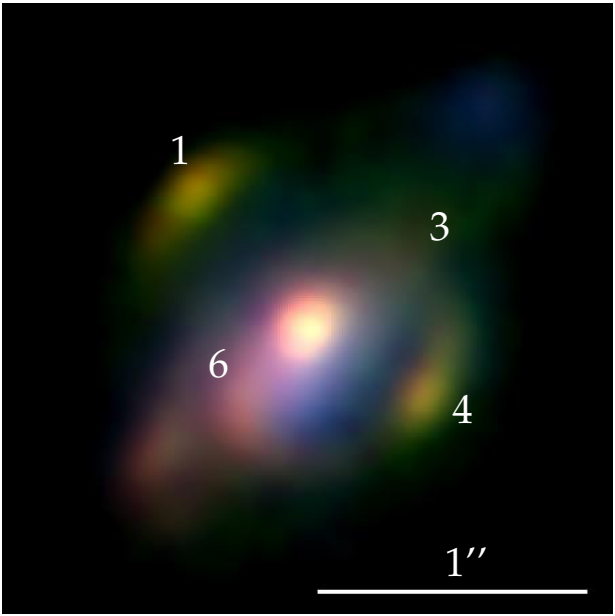}
\caption{\label{fig:colour} Color image from a RGB composite of the
  WFPC2 F814W, NICMOS F160W and NIRC2 Kp images.  The lensed arcs
  correspond to radio image components 1, 3, 4, and 6 of the compact
  core.  Component 3 is barely visible and component 6 is reddened due
  to dust extinction in the plane of the spiral galaxy.} 
\end{figure}

Archival \HST\ images of \ourlens\ in the F160W filter with the Near
Infrared Camera and Multi-Object Spectrometer (NICMOS; Proposal ID:
9744; PI: Kochanek) and the F555W and F814W filters with the Wide
Field and Planetary Camera 2 (WFPC2; Proposal ID: 9133; PI:
Falco) are available.  We used {\sc MultiDrizzle}\footnote{{\sc
    MultiDrizzle} is a product of the Space Telescope Science
  Institute, which is operated by AURA for NASA.} 
to combine the exposures in each filter and correct for
geometric distortion.   The F555W data were discarded due to the low
signal-to-noise ratio (SNR).  We
oversampled the F814W and F160W images with 
{\sc MultiDrizzle} to the same pixel scale as the
NIRC2 image, and show the color image composed of the three filters in
Figure \ref{fig:colour}.  The images of the lensed source in the plane
of the lens galaxy (corresponding to radio components 3 and 6) suffer
from dust extinction.  

We use the stars in the WFPC2 F814W images to obtain the pixel scale of the
NIRC2 image in Section \ref{sec:obs:AO} and the alignment of radio and
NIRC2 images in Section \ref{sec:align}.  Furthermore, we use the F814W
and F160W photometries in Section \ref{sec:lenslight:phot} to estimate
the stellar mass of the lens 
galaxy (that incorporates the effects of dust) in Section \ref{sec:IMF}.

\section{Radio and NIRC2 image alignment}
\label{sec:align}
In order to use both the radio image positions of the source and the
NIRC2 image of the lens galaxy to constrain the lens mass distribution, we need
to align the radio and the NIRC2 images.
We assume that the two coordinate systems differ by a rotation and a
translation.  
By aligning the stars in the NIRC2 images to the corresponding stars in
the WFPC2 images with WCS information, we determine the north
direction on the NIRC2 image, and consequently, the rotation between the
radio and NIRC2 images.  To determine the translational offset between
the radio and NIRC2 images, we use the centroid positions of the two
prominent arcs in the NIRC2 image which we denote as $P_1$ and $P_2$.
The separation between $P_1$ and $P_2$ and the angle of the segment
connecting $P_1$ and $P_2$ match those of components 1 and 4 within
$5$\,mas and $0.3\degr$, respectively.  The matching of $P_1$ and
$P_2$ to components 1 and 4 agrees with previous identifications of
features in the F160W image of \citet{Marlow++99}.  We construct the
coordinate translation from the NIRC2 to the radio image by mapping the
midpoint of the segment connecting $P_1$ and $P_2$ to the midpoint of
the segment connecting components $1$ and $4$.  We show in Figure
\ref{fig:AOimage} the superposition of the radio and NIRC2 data sets.

The positions $P_1$ and $P_2$ of the arcs in the Keck infrared
image can be determined with an accuracy of $5\,$mas. 
We add this systematic alignment uncertainty to the
positional uncertainty of the galaxy centroid in the mass modeling.
We also incorporate the uncertainty in the rotation between the
NIRC2 and radio image ($0.3\degr$) to the uncertainty in the PA 
of the lens galaxy. 

Having determined the transformation from the NIRC2 image to the
radio observations, all coordinates
values are reported with respect to the system used in
Table \ref{tab:image-positions} for the remainder of the paper.

\section{Lens galaxy light distribution}
\label{sec:lenslight}

In this section we describe the steps taken to model the light profile
of the lens galaxy in the NIRC2 image.

\subsection{PSF}
A model of the point spread function (PSF) is needed to extract
the intrinsic light profile of the lens galaxy without atmospheric and
instrumental blurring.  We use the
star that is $\sim$$2\farcs9$ northwest of \ourlens\ in the observed field
as a model of the PSF.
Previous works have shown that field stars serve as good PSF models,
especially for spatially extended objects \citep[e.g.,][]{Marshall++07,
  Suyu++09}.

\subsection{Lens light profile}
\label{sec:lenslight:profile}

To model the lens galaxy light, we use the {\sc Galfit}
package \citep{Peng++02}.  We mask out the
lensing arcs and the spiral arm-like features in the
southeast region of 
the galaxy. Optionally, we also mask out the central region of the
galaxy.  
We find that the galaxy is well described by two exponential
disk profiles with a common centroid (or a single exponential disk
profile, if the central region was masked out).  Specifically, we employ
S{\'e}rsic profiles with index $n_{\rm sersic}\equiv 1$ that correspond to 
exponential disks and also fit a uniform background for the sky.  One of the S{\'e}rsic profiles corresponds to the
disk of the galaxy, and the other is centrally localized with a small
effective radius ($\sim$$0\farcs05$)\footnote{The effective
    radius of the central component remains small
  ($<0\farcs07$) even when the S{\'e}rsic index is allowed to
  vary between 1 and 4.}.  We identify this latter component
as a bulge, although it could also be a bar given the limited resolution.
We estimate the uncertainties on the S{\'e}rsic parameters based on differences
in the best-fit parameter values for different choices of masks. This
systematic uncertainty dominates the statistical uncertainty of the fit. 
For the galaxy centroid, we include the systematic uncertainty of 5\,mas
(from the alignment of NIRC2 and radio data) which dominates the overall
positional uncertainty.
Table~\ref{tab:best-fit-light} lists the best-fit values for the
mask containing the lensed arcs and spiral arm features.

\begin{table*}[t]
\caption{Lens galaxy light}
\label{tab:best-fit-light}
\begin{center}
\begin{tabular}[b]{lcccccc}
\hline
& $\Delta$R.A. & $\Delta$Dec & $R_{\rm e}$ & $n_{\rm sersic}$ & $q$ & $\phi$ \\

& (arcsec) & (arcsec) & (arcsec) & & & ($\degr$) \\\hline 
%Masked arcs and features in coordinate of Cohn et al.
Disk  & $0.040\pm0.005$ & $-0.036\pm0.005$ & $0.85\pm0.05$ & $\equiv$$1$ & $0.63\pm 0.06$ & $138.0\pm1.5$ \\
Bulge  & $0.040\pm0.005$ & $-0.036\pm0.005$ & $0.055\pm0.006$  & $\equiv$$1$ & $0.41\pm0.02$ & $147\pm3$ \\\hline
\end{tabular}
\end{center}
Notes.  S{\'e}rsic profile parameters for the galaxy disk and bulge
based on the NIRC2 image.  Columns 2 and 3 are the relative
right ascension and declination, respectively, in the coordinate system of
\citet{Cohn++01} where the origin is close to the lens center.
Column 4, 5, 6 and 7 are the effective radius, S{\'e}rsic index, axis
ratio and position angle of the S{\'e}rsic profile, respectively.
\end{table*}

\begin{figure*}
\centering
\includegraphics[width=\textwidth]{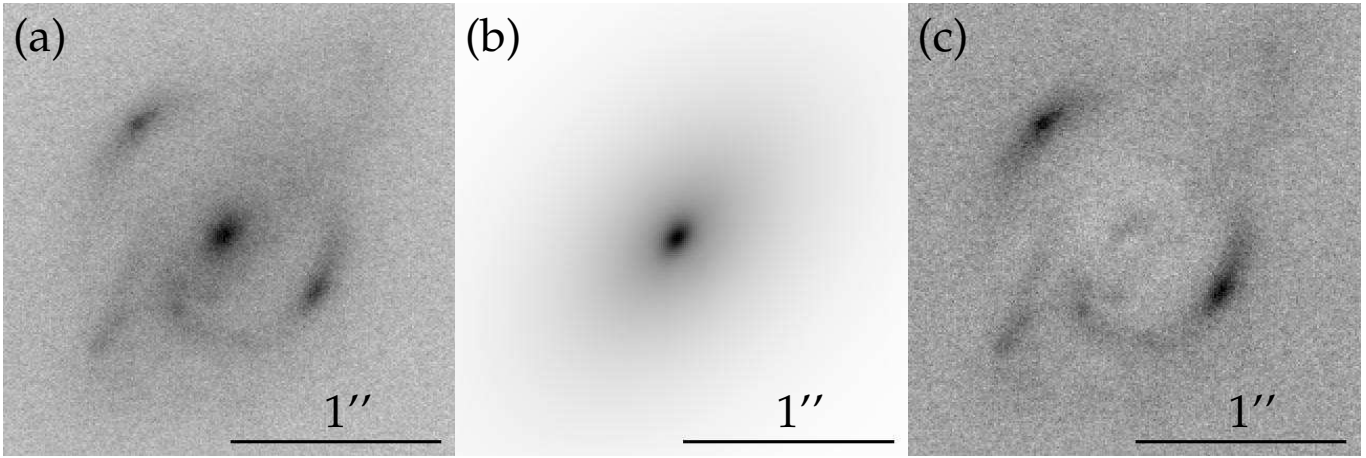}
\caption{Lens galaxy light. (a) NIRC2 Kp image, (b) modeled lens
  galaxy light based on two exponential disk profiles for the galaxy disk
  and bulge, (c) residual image. }
\label{fig:galfit}
\end{figure*}

We see in Figure \ref{fig:galfit} that the two-component S{\'e}rsic model
reproduces the overall structure of the lens galaxy light, and
  the background fit yields uniform sky residuals. The reduced $\chi^2$ is
  1.03 in the fitting region.
The
residuals show, apart from the three strong lensing arcs
(corresponding to radio components 1, 4, and 6 in Figure \ref{fig:AOimage}),
some small-scale features that could be spiral arms or tidal features
in the lens galaxy.  In Sections \ref{sec:kinematics:results} and
\ref{sec:lensing:results}, we account 
for the effects of these residual features in the mass modeling by
adding systematic uncertainties to the line-of-sight velocities and
the radio image positions.

\subsection{Lens photometry}
\label{sec:lenslight:phot}

The photometry of the lens is required for estimating the stellar mass
from SPS models in Section \ref{sec:IMF}.  
To obtain the integrated K-band magnitudes for the disk and the bulge
based on the exponential profiles in the previous section,
we first determine the photometric zero point of the NIRC2 image by 
calibrating it with the low resolution NIRC K-band image. We then
integrate the model light profiles and list in Table
\ref{tab:lensmag} the K-band magnitudes of the disk and the bulge,
where the estimated uncertainty comes from calibration, difference
in throughput of NIRC2 Kp and NIRC K filters, and variations
between different arc masks.  
 
The \HST\ images have significantly lower SNR than the NIRC2 image.
Therefore, we use the structural parameters (axis ratio, position
angle, and effective radius) of the exponential profiles from the NIRC2
image (listed in Table \ref{tab:best-fit-light}) to obtain the
integrated magnitudes in F814W and F160W with {\sc Galfit}.  We use the
nearest field star to approximate the PSF.  The separation of the arcs and
the lens light is more difficult in these low SNR images, and we
conservatively adopt an uncertainty of 0.3 magnitudes from various
choices in the arc masks.  The integrated magnitudes for the disk and the bulge
are listed in Table \ref{tab:lensmag}.

\begin{table}
\caption{Lens Photometry}
\label{tab:lensmag}
\begin{center}
\begin{tabular}{lccc}
\hline
      & WFPC2 F814W  & NICMOS F160W & NIRC2 K \\
\hline
Disk  & $19.0\pm0.3$ & $17.5\pm0.3$ & $18.5\pm0.3$ \\ 
Bulge & $23.0\pm0.3$ & $22.1\pm0.3$ & $22.9\pm0.3$ \\
\hline
\end{tabular}
\end{center}
Notes. Magnitudes are in the AB system.  
\\
\end{table}

\section{Galaxy mass components}
\label{sec:mass-model}

We decompose the spiral galaxy into three mass components: a 
disk of stars and gas, a bulge and a dark matter halo.
We now briefly describe the mass distribution
for each component.

\subsection{Disk}
We assume that the disk of stars and gas in the galaxy is thin and
circularly symmetric.
Furthermore, we assume that there is a constant $M/L$
throughout the disk.  Hence, the exponential profile for the
light in Section \ref{sec:lenslight:profile} implies that the profile for the
elliptical surface mass density of the projected (inclined) disk is
\be
\label{eq:diskSigma}
\Sigma_{\rm d}^{\rm P}(R') = \Sigma_{\rm d,0}\,\exp\left[-R'/R_{\rm d} \right],
\ee
where $\Sigma_{\rm d,0}$ is the normalization of the disk (set by the
$M/L$), $R_{\rm d}$ is the scale radius of the disk, and $R'$ is given in
terms of the coordinates $x',y'$ along the major/minor axes centered 
on the galaxy by
\be
R'^2 = x'^2 + (y'/q_{\rm d})^2. 
\ee
The axis ratio $q_{\rm d}$ is related to the inclination angle $i$ of the
disk by 
\be
q_{\rm d}=\cos i,
\ee
where $i=0\degr$ corresponds to a face-on disk.

In terms of the radial coordinates in the plane of the disk, $R=\sqrt{x^2+y^2}$, the surface mass density is 
\be
\label{eq:diskFrameSigma}
\Sigma_{\rm d}(R) = \Sigma_{\rm d,0}\,q_{\rm d}\,\exp\left[
  -R/R_{\rm d}\right],
\ee
where the extra factor of $q_{\rm d}$ in the normalization ensures that an
inclined disk and a face-on disk have the same total mass.  The
scale radius $R_{\rm d}$ is related to the effective radius (the
half-light radius in Table \ref{tab:best-fit-light}, which is also
the half-mass radius with a constant $M/L$) by $R_{\rm d}=R_{\rm
  e}/1.678$.  

For gravitational lensing, the quantity of interest is the
dimensionless surface mass density,
\bea
\kappa_{\rm d}(R')&=&\Sigma_{\rm d}^{\rm P}(R')/\Sigma_{\rm crit}\\
 &=& \kappa_{\rm d,0}\,\exp\left[-R'/R_{\rm d} \right],
\eea
where 
\be
\Sigma_{\rm crit} = \frac{c_{\rm l}^2}{4\pi G}\frac{D_{\rm s}}{D_{\rm d}
  D_{\rm ds}},
\ee
$\kappa_{\rm d,0}$ is the disk strength, $c_{\rm l}$ is the speed of
light and $G$ is the gravitational constant. The 
distances $D_{\rm d}$, $D_{\rm s}$, and $D_{\rm ds}$ are angular
diameter distances to the lens, to the source, and between the lens
and the source, respectively.

\subsection{Bulge}
The galaxy light fitting suggests that we can model the bulge as a
point mass, since the effective radius of the corresponding
S{\'e}rsic profile is very small compared to the radial range of
positions spanned by the lensed images ($\sim$$0\farcs3$ to $0\farcs8$) and
most of the points in the rotation curve.

\subsection{Dark matter halo}
We assume a NFW profile \citep{Navarro++96} with a triaxial shape
\citep{JingSuto02} for the dark matter halo.  The three-dimensional
density distribution is given by
\begin{equation}
\label{eq:tnfw}
\rho_{\rm h}(r) = \frac{\rho_{\rm h,0}}{(r/r_{\rm
    h,0})(1+r/r_{\rm h,0})^2}
\end{equation}
where
\begin{equation}
\label{eq:tnfw_r}
  r^2 = c^2\left(\frac{x^2}{a^2} + \frac{y^2}{b^2} +
    \frac{z^2}{c^2}\right), \quad a\leq b \leq c.
\end{equation}

The parameters $a, b$, and $c$ describe the triaxial shape of the
halo.  The orientation of the dark matter halo as seen by a distant
observer can be described by two angles $\vartheta$ and $\varphi$.
Following \citet{Oguri++03}, we choose ($\vartheta$,$\varphi$) as the
polar angle of the observer's line-of-sight direction in the halo
coordinate system $(x,y,z)$ (see Figure 1 of \citet{Oguri++03}).  With
this definition, equations (1) and (2) of \citet{Oguri++03} relate
the coordinates in the frame of the halo $(x,y,z)$ to the coordinates
of the distant observer $(x',y',z')$.  

The two dimensional surface mass density of the projected triaxial
halo is elliptical, and the axis ratio ($q_{\rm h}$) and position
angle ($\phi_{\rm h}$) of the ellipse can be calculated given the
values for $a/c$, $b/c$, $\vartheta$ and $\varphi$ \citep{Oguri++03}.
We use the Einstein radius for the corresponding spherical mass
model, $\RhE$, to characterize the strength of 
the halo since lensing can robustly measure this quantity.  The
Einstein radius is the radius of the ring that is formed by a point source
lying perfectly behind a spherical halo with strength $\RhE$.  The
normalization $\rho_{\rm h,0}$ is related to the Einstein radius
$\RhE$ by
\be
\rho_{\rm h,0} = A \left\{ \begin{array}{ll}
\left[1-\log(2) \right]^{-1} & \rm{if\ } \tRhE=1 \phantom{\frac{\Sigma}{\Sigma}}  \\
\Big{[} \log\left(\frac{\tRhE}{2}\right) +
  \frac{2}{\sqrt{1-\tRhE^2}}\cdot & \\
  \ \ \ \rm{arctanh} \left(\sqrt{\frac{1-\tRhE}{1+\tRhE}}\right)
  \Big{]}^{-1} & \rm{if\ } \tRhE<1\\
\Big{[} \log\left(\frac{\tRhE}{2}\right) +
  \frac{2}{\sqrt{\tRhE^2-1}}\cdot & \\
  \ \ \ \rm{arctan} \left(\sqrt{\frac{\tRhE-1}{\tRhE+1}}\right)
  \Big{]}^{-1} & \rm{if\ } \tRhE>1\\

\end{array} \right.
\ee
where
\be
A=\frac{\Sigma_{\rm crit}}{4 r_{\rm h,0} D_{\rm d}} \tRhE^2,
\ee
and $\tilde{R}_{\rm h,E}=\RhE/r_{\rm h,0}$.  Both $R_{\rm
  h,E}$ and $r_{\rm h,0}$ are in arcseconds (which can be easily
converted to, e.g., kpc, with the angular diameter distance to the
lens, $D_{\rm d}$).

\subsection{External shear}
For the lensing analysis in Section \ref{sec:lensing}, we also
include an external shear component with strength $\gamma_{\rm ext}$
and position angle $\phi_{\rm ext}$.  A shear angle of $\phi_{\rm
  ext}=0$ corresponds to an elongation of the images in the east-west
direction.  

\subsection{Combined mass model}
\label{sec:mass-model:comb}

For simplicity, we impose a 
symmetry condition on the model. To be precise, we require that the
centroids of 
the three components coincide and that the total mass distribution is
three-dimensionally axisymmetric. 
In particular, the NFW profile is either prolate ($a=b$) or oblate
($b=c$).  Based on the symmetry assumptions and the axis ratio and PA
of the projected disk (in Table \ref{tab:best-fit-light}), the
orientation of the halo is
$(\vartheta,\varphi)=(50\pm5\degr,145\pm3\degr)$.  

\citet{Cohn++01} found that the orientation of the quadrupole moment of
their lens model (one-component total mass profile and external shear)
agrees with the position angle of the lens galaxy within
$\sim$$5\degr$.  This implies that the position angle of the projected total
mass distribution is aligned with the light distribution of the
galaxy.  We confirm the alignment by modeling the lens system using a
pseudoisothermal elliptic mass distribution 
\citep{KassiolaKovner93} in the presence of external shear.  The
alignment of the projected total mass distribution and the light implies that
the NFW halo is oblate (in the inner region probed by lensing) since a
prolate halo would lead to a 
$\sim$$90\degr$ difference in the position angles of the projected
mass and of the light distributions. 

We impose Gaussian priors on (i) the centroid of the total
  mass distribution, (ii) the projected axis ratio and scale 
radius of the disk, and (iii) the orientation of the NFW halo, based
on the observed 
light profile in Table \ref{tab:best-fit-light}.  The position angle
of the projected disk is set by the orientation of the NFW halo (based
on the axisymmetry 
assumption).  We impose uniform priors on the remaining
  parameters, and summarize the priors in Table \ref{tab:priors}.
While all these priors
are imposed for the lens modeling, some are not needed for the
kinematics, such as the external shear\footnote{\ourlens\ is not in
  any obvious galaxy group or cluster, so any external shear is likely to come
  from structures that are not dynamically associated with the lens
  system.}.  
In total, we have 13 mass parameters: 6 with
  Gaussian priors, and 7 ``free'' parameters with uniform priors.

\begin{table}
\caption{Priors on model parameters}
\label{tab:priors}
\begin{center}
\begin{tabular}{l l l}
\hline
Description & Parameter$\phantom{00}$ & Prior \\
\hline
Centroid in $\theta_1$ & $\theta_{\rm 1,c}$ & $G(-0\farcs040, 0\farcs005)$ \\
Centroid in $\theta_2$ & $\theta_{\rm 2,c}$ & $G(-0\farcs036, 0\farcs005)$ \\
Disk axis ratio & $q_{\rm d}$ & $G(0.63,0.06)$ \\
Disk strength & $\kappa_{\rm d,0}$ & $U(0,\infty)$ \\
Disk scale radius & $R_{\rm d}$ & $G(0\farcs51, 0\farcs03)$ \\
Bulge Mass & $M_{\rm b}$ & $U(0,\infty)$ \\
Halo flattening & $a/c=a/b$ & $U(0.25,1)$ \\
Halo orientation angle & $\vartheta$ & $G(50\degr, 5\degr)$ \\
Halo orientation angle & $\varphi$ & $G(145\degr, 3\degr)$ \\
Halo Einstein radius & $\RhE$ & $U(0,\infty)$ \\  
Halo scale radius & $r_{\rm h,0}$ & $U(0\farcs1,8'')\equiv U(0.75,60)\kpc$ \\
External shear strength & $\gamma_{\rm ext}$ & $U(0,0.3)$ \\
External shear angle & $\phi_{\rm ext}$ & $U(0,2\pi)$ \\
\hline
\end{tabular}
\end{center}
Notes. $G(\mu,\sigma)$ is a Gaussian distribution with mean $\mu$
and standard devation $\sigma$.  $U(a,b)$ is a uniform distribution
between boundaries $a$ and $b$.  In cases where $b=\infty$, the upper
boundary is set to a real number that corresponds 
to masses $\gtrsim 10^{13}\,{\rm M}_{\sun}$, i.e., larger than
galaxy-scale masses. 

\end{table}

\section{Bayesian Analysis}
\label{sec:Bayes}
We use Bayesian analysis to infer the mass model parameters.  In
particular, we sample the posterior probability distribution function
(PDF) of the 13 mass parameters $\paramvec$ which is given by
\be
\label{eq:bayes}
P(\paramvec|\data) =
\frac{\overbrace{P(\data|\paramvec)}^{\rm likelihood}\,\overbrace{P(\paramvec)}^{\rm
    prior}}{\underbrace{P(\data)}_{\rm evidence}}
\ee
where $\data$ is the data.  The expressions for the likelihoods of
the kinematics and the lensing data are in Sections
{\ref{sec:kinematics:results}} and {\ref{sec:lensing:results}},
respectively.  The prior $P(\paramvec)$ is given by 
\be
\label{eq:prior}
P(\paramvec)=\prod_{i=1}^{13} P(\param_i),
\ee
where $\param_i$ is the $i$-th parameter, and $P(\param_i)$ is either
a Gaussian or a uniform distribution as described in Section
\ref{sec:mass-model:comb}.  The Bayesian evidence is used for
comparing different forms of models, which does not concern us in this
paper since we consider only one form of mass model: an exponential disk, a
point mass bulge and an oblate NFW halo.

\section{Rotation curve modeling}
\label{sec:kinematics}

In this section we describe the modeling of the galaxy mass
distribution based on the rotation curve constraints.
We assume that the gas in the disk (from which we observed the emission
lines for the rotation curve) is in circular orbits, and use the circular
velocities of the mass model to predict the line-of-sight velocities
for constructing the likelihood of the kinematics data.  We then sample
the posterior PDF of the kinematics data.

\subsection{Circular velocities}

For a test mass in the plane of the disk at a radius $R$
from the center, its circular velocity $v_{\rm c}(R)$ has three
contributions: 
\be
v_{\rm c}^2 = v_{\rm b}^2 + v_{\rm d}^2 + v_{\rm h}^2,
\ee
where $v_{\rm b}, v_{\rm d}$, and $v_{\rm h}$ are the circular velocities
of the bulge, disk and halo mass distributions, respectively.

For the point-mass bulge, the rotational velocity
contribution $v_{\rm b}$ takes on a simple form
\be
v_{\rm b}^2 = \frac{G M_{\rm b}}{R},
\ee
where $G$ is the gravitational constant and $M_{\rm b}$ is the mass of
the bulge. 

The contribution of the disk also has an analytic expression
in terms of Bessel functions $I_0$, $K_0$, $I_1$, and $K_1$ 
\citep[e.g.,][]{BinneyTremaine87}:
\be
v_{\rm d}^2 = 4\pi G\Sigma_{\rm c} R_{\rm d} y^2 (I_0(y)K_0(y) -
I_1(y)K_1(y)),
\ee
where $\Sigma_{\rm c}$ is the central surface mass density of the
disk ($\equiv \kappa_{\rm d,0} \Sigma_{\rm crit} q_{\rm d}$), $R_{\rm
  d}$ is the scale radius of the disk, 
$y$ is related to the radial distance $R$ by $y =
\frac{R}{2R_{\rm e}}$, and $R_{\rm e}$ is the effective 
radius of the mass profile of the disk.

For the halo component, we follow \citet{BinneyTremaine87} and
integrate the oblate spheroidal three-dimensional mass density
$\rho(m)$ with $m^2=x^2/q_{\rm h}^2+y^2+z^2$ to obtain the
circular velocity: 
\be
\label{eq:vhalo}
v_{\rm h}^2 = 4\pi G q_{\rm h} \int_0^{R}
m^2\frac{\rho(m)}{\sqrt{R^2-m^2(1-q_{\rm h}^2)}}{\rm d}m,
\ee
where $q_{\rm h} = a/c$ is the flattening of the dark matter halo.
For the NFW mass density (equation (\ref{eq:tnfw})), we numerically
compute the integral in equation (\ref{eq:vhalo}).

\subsection{Predicted rotation curve}
To model the rotation curve, we follow \citet{vanderMarelvanDokkum07}
to obtain the predicted line-of-sight velocities from the circular
velocities which incorporate the effects of the disk inclination, 
slit and seeing.

The line-of-sight velocity in the absence of the slit and
seeing at a point $(x',y')$ on the sky (with $(x',y')=(0,0)$ at the center 
of the galaxy and $x'$ along the major axis of the lens galaxy) is
\be
\label{eq:losv}
v_{\rm los}(x',y') = \frac{x}{R} \sin(i)
v_{\rm c}(R),
\ee
where $\boldsymbol{R}=(x, y)$ is the
corresponding coordinate in the plane of the disk with $x=x'$,
$y=\frac{y'}{\cos i}$, and $R=\vert \boldsymbol{R} \vert$.  

We weight $v_{\rm los}$ by the modeled galaxy light, and convolve with
both a Gaussian PSF with FWHM of $0\farcs6$ and a square top-hat function
of size $W \times P$ (for slit of width $W$ and pixel scale $P$) for
each (binned) spatial pixel along the slit.  The resulting weighted
and convolved velocity is the predicted line-of-sight velocity
$v_{\rm los}^{\rm pred}$ at the corresponding spatial pixel.

\subsection{Kinematics analysis and results}
\label{sec:kinematics:results}
The uncertainties of the data points in the rotation curve of Figure
\ref{fig:rotcurve} are only statistical.  We include a systematic
uncertainty of $40\kms$ (which is comparable to the statistical
uncertainty) to account for deviations from our assumptions 
of the smooth axisymmetric mass model, slight offset (if any) of the
galaxy centroid in the rotation curve, and the unknown systemic velocity. 
We explore a range of values for the systemic velocity, and
choose the value that optimizes the rotation curve fitting.  We
add the statistical and the 
systematic uncertainty in quadrature to obtain the final uncertainty
on the line-of-sight velocity.  The amount of systematic uncertainty
is set such that the rotation curve can be modeled with a reduced
$\chi^2$$\sim$$1$.  The inclusion of the systematic 
uncertainty is crucial for not underestimating the uncertainty on the 
resulting mass model parameters.

The likelihood for the rotation curve data is
\be
\label{eq:dynlike}
P(\data_{\rm D} | \paramvec) = \frac{1}{Z_{\rm D}} \exp \left[ -\frac{1}{2}
  \sum_{i=1}^{N_{\rm D}} \frac{\left( v_{{\rm los},i}^{\rm obs} - v_{{\rm
        los},i}^{\rm pred} \right)^2} {\sigma_i^2} \right],
\ee
where $N_{\rm D}$ is the number of data points ($=15$), 
$v_{{\rm los},i}^{\rm obs}$ is the observed line-of-sight
velocity of data point $i$, $v_{{\rm los},i}^{\rm pred}$ is the
predicted line-of-sight velocity from our model and observational setup, 
$\sigma_i$ is the uncertainty in the velocity, and $Z_{\rm D}$
is the normalization given by
\be
Z_{\rm D} = (2\pi)^{N_{\rm D}/2} \prod_{i=1}^{N_{\rm D}} \sigma_i.
\ee

\begin{figure}
  \centering
  \includegraphics[width=0.52\textwidth]{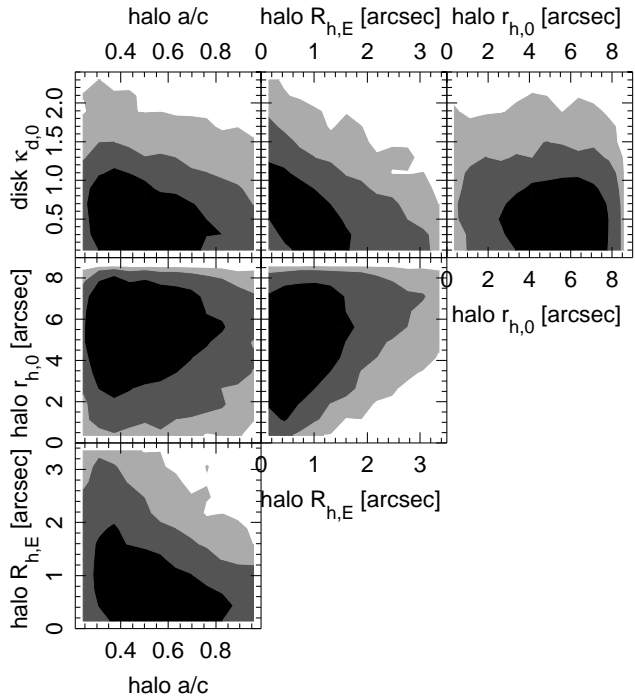}
  \caption{ \label{fig:PDF_Dyn} Marginalized posterior PDF for
    the halo flattening $a/c$, halo Einstein radius $\RhE$, halo
    scale radius $r_{\rm h,0}$, and disk strength $\kappa_{\rm d,0}$ based on
    only the rotation curve data.  The three shaded areas show the 68.3\%,
    95.4\% and 99.7\% credible regions.  The disk-halo degeneracy is
    illustrated in the top-middle panel: higher $\RhE$ corresponds
    to lower $\kappa_{\rm d,0}$.}
\end{figure}

We use {\sc MultiNest} \citep{FerozHobson08, Feroz++09} to sample the posterior PDF of the
13 model parameters.  Figure~\ref{fig:PDF_Dyn} shows the resulting
constraints  on four of 
the parameters: halo flattening $a/c$, halo Einstein radius $\RhE$,
halo scale radius 
$r_{\rm  h,0}$, and disk strength 
$\kappa_{\rm d,0}$.  The disk-halo degeneracy appears as the
anti-correlation between the disk strength ($\kappa_{\rm d,0}$) and
the halo Einstein radius ($\RhE$): the more massive the disk, the less
massive the halo, and vice versa.  The halo scale radius is
poorly constrained.  Furthermore, the kinematics data provide little 
information on the halo shape.

\begin{figure}
\centering
\includegraphics[width=0.36\textwidth, angle=270]{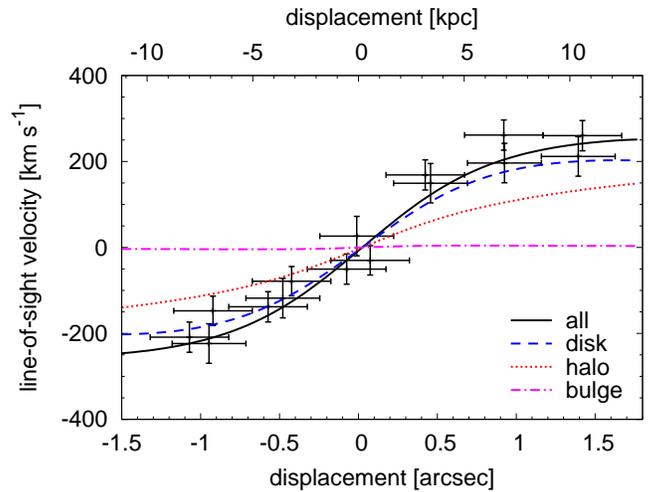}
\caption{ \label{fig:MProtcurv} Rotation curve of the most probable
  mass model based on kinematics data only.  The contributions
  from the disk (blue dashed), halo (red dotted) and bulge (magenta
  dot-dashed) are indicated. Most of the mass is in the disk and
  the dark matter halo.}
\vspace{1cm}
\end{figure}

Figure~\ref{fig:MProtcurv} shows the predicted rotation curve of the
most probable mass distribution with a reduced $\chi^2=1.0$ based on
the kinematics data.  The 
bulge contributes very little to the mass of the system, consistent
with the small effective radius observed for the bulge light
distribution.  At small (large) radii, the disk (dark matter halo)
dominates in the enclosed mass.

\section{Lens modeling}
\label{sec:lensing}
In this section we discuss the properties of our three-component mass
model based on the lensing constraints.

\subsection{Lensing deflection angles}
We briefly describe how to obtain the 
deflection angles for the three mass components in our model.
The deflection angles are used to solve for the predicted radio image
positions that are needed for constructing the likelihood of the
lensing data.

\subsubsection{Disk}
The lensing deflection angles for an elliptical mass component
following an exponential disk 
profile has no analytical form. Hence, to model the stellar
disk component, we employ a ``chameleon profile'' 
mimicking the mass profile of an exponential disk whose deflection
angles are analytical.  This profile is inspired by the chameleon
profile used in \citet{Maller++00}.  
We use the following approximation for the
dimensionless surface mass density of the disk:
\bea
\kappa_{\rm d}(R') & = & \kappa_{\rm d,0} \exp\left[-R'/R_{\rm d}\right] \\
          \label{eq:chamdisk}
          & \approx & \kappa_{\rm d,0} s_1 \left[
            \frac{1}{\sqrt{R'^2+s_2^2}} -
            \frac{1}{\sqrt{R'^2+s_3^2}}\right], 
\eea
where $s_1=4.6849R_{\rm d}$, $s_2=1.1720R_{\rm d}$ and
$s_3=1.4518R_{\rm d}$.  The
projected enclosed mass within radius $\tilde{R'}$ for the chameleon profile
agrees with that of the exponential profile within $2\%$ for the
range of radii spanned by the lensed images ($\tilde{R'}=0.6R_{\rm d}$ to
$1.6R_{\rm d}$).
Each term in the square brackets in equation (\ref{eq:chamdisk}) is in
the form of a pseudoisothermal elliptic mass distribution whose
deflection angles can be computed analytically
\citep{KassiolaKovner93}.

\subsubsection{Bulge}
The bulge is modeled as a point mass, and the deflection angle
has a simple closed form \citep[e.g.,][]{Schneider++06}
\begin{equation}
  \alpha(\xi) = \frac{4G}{c_{\rm l}^2}\frac{M_{\rm b}}{\xi}
\end{equation}
where $G$ is the gravitational constant, $c_{\rm l}$ is the speed of light,
$M_{\rm b}$ is the mass of the bulge, and $\xi$ is the impact parameter
(i.e., the distance between the light ray and the point mass on the
lens plane).

\subsubsection{Dark matter halo}
We refer the reader to \citet{Oguri++03} for the lensing properties of
the triaxial NFW halo in equations (\ref{eq:tnfw}) and
(\ref{eq:tnfw_r}).  The projected surface mass density of the halo is
elliptical, and the deflection angles can be computed numerically.

\subsection{Lensing analysis and results}
\label{sec:lensing:results}

We use the radio image positions listed in Table
\ref{tab:image-positions} to constrain the mass model parameters.
We add a positional uncertainty of $10$\,mas in quadrature to the
uncertainties in the table to account for systematic effects such as
the residual features in the lens galaxy light fit
(Section \ref{sec:lenslight:profile}), presence of substructure
\citep[e.g.,][]{Chen++07} and scatter-broadening through the disk of the lens
\citep[e.g.,]{Marlow++99, Biggs++04}.  The lensing arcs in
the NIRC2 images could in principle be used in addition to the radio
image positions to constrain the lens mass distribution; however,
these arcs (especially the one associated with components 3 and 6) in
practice suffer from dust extinction.  Only the F160W image of the
\HST\ data has sufficient SNR to be used for dust
correction, and \citet{Suyu++09} showed that dust correction based on
two bands are prone to systematic effects.  Therefore, we do not
include the dust-affected NIRC2 arcs for constraining the mass
distribution.

The likelihood of the radio image positions is
\be
\label{eq:lenslike}
P(\data_{\rm L} |\paramvec) = \frac{1}{Z_{\rm {L}}} \exp
  {\left[-\frac{1}{2}\displaystyle\sum_{j=1}^{N_{\rm
          sys}}\displaystyle\sum_{i=1}^{N_{\rm im}^j}
      \frac{\vert\boldsymbol{\theta}_{i,j}^{\rm obs}-\boldsymbol{\theta}_{i,j}^{\rm
          pred}(\paramvec)\vert^2}{\sigma_{i,j}^2} \right]},
\ee
where $N_{\rm sys}$ is the number of multiply imaged systems, 
$N_{\rm im}^j$ is the number of multiple images in system
$j$, $\boldsymbol{\theta}_{i,j}^{\rm obs}$ is the observed image position, 
$\boldsymbol{\theta}_{i,j}^{\rm pred}(\paramvec)$ is the modeled image
position, $\sigma_{i,j}$ is the uncertainty in the observed image
position, and $Z_{\rm L}$ is the normalization given by
\be
\label{eq:lenslike_norm}
Z_{\rm L}={(2\pi)^{N_{\rm L}/2}
  \displaystyle\prod_{j=1}^{N_{\rm sys}} \displaystyle\prod_{i=1}^{N_{\rm im}^j} \sigma_{i,j}}
\ee
with
\be
N_{\rm L}=\displaystyle\sum_{j=1}^{N_{\rm sys}} {N_{\rm im}^j}.
\ee
The source position for each system of multiple images is needed to
predict the image positions.  We model the source position as the
weighted average of the
mapped source positions from the observed image positions and the
deflection angles from the mass distribution. Specifically, for each
image system, we take the average of the mapped source position
$\boldsymbol{\beta_i}$ weighted by $\sqrt{\mu_{i}}/\sigma_{i}$, where
$\mu_{i}$ and $\sigma_{i}$ are, respectively, the modeled
magnification and the positional uncertainty of image
$i$.\footnote{Our tests using analytic mass profiles show that the
  weighted source positions and the optimized source positions give
  consistent constraints on the mass parameters.  Since the
  computation time of the  
weighted source positions is an order of magnitude shorter than that
of the optimized source positions and the deflection angles of the
oblate NFW halo are computationally expensive (due to the numerical
integration), we use the weighted source positions.}  In doing so, we 
have in effect  
marginalized the source position parameters by approximating the
lensing likelihood as having a delta function at the weighted source
position for each image system.

We model the 3-component mass distribution of the spiral galaxy based
on the lensing data with {\sc Glee} (``Gravitational Lens Efficient
Explorer''), a software developed by A.~Halkola \&
S.~H.~Suyu\footnote{The software models the mass distribution in
  gravitational lenses using image positions \citep{Halkola++08} or
  extended images with pixelated source \citep{Suyu++06}.  We refer
  the reader to \citet{SuyuHalkola10} for an example of each
  approach.}.  For a set of mass model parameter values, {\sc Glee} computes
the values of the lensing likelihood in equation (\ref{eq:lenslike})
and the prior in equation (\ref{eq:prior}); these are the two
ingredients needed to obtain the posterior PDF in equation
(\ref{eq:bayes}).  As in the kinematics analysis, we use {\sc MultiNest}
\citep{FerozHobson08, Feroz++09} to sample the posterior PDF.

\begin{figure}
  \centering
  \includegraphics[width=0.52\textwidth]{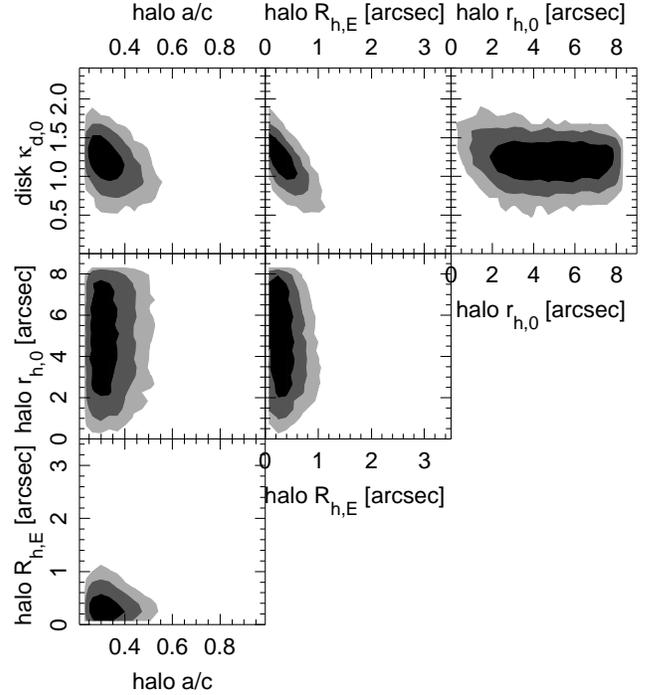}
  \caption{ \label{fig:PDF_Lens} Marginalized posterior PDF for
    the halo flattening $a/c$, halo Einstein radius $\RhE$, halo
    scale radius $r_{\rm h,0}$, and disk strength $\kappa_{\rm d,0}$ based on
    only the lensing data.  The three shaded areas show the 68.3\%,
    95.4\% and 99.7\% credible regions.  The panels are plotted on the
    same scales as in Figure \ref{fig:PDF_Dyn} for comparison.}
\end{figure}

Figure~\ref{fig:PDF_Lens} shows the resulting constraints on the same
parameters as those in Figure \ref{fig:PDF_Dyn}.  The degeneracy
between the disk strength and the Einstein radius of the dark matter
halo is visible, though not as strong as in the case of the
kinematics-only analysis.  The halo is highly flattened ($a/c\,\sim\,0.3$).
The flattening is degenerate with the disk strength as shown in the
top-left panel: a flattened halo is less massive and requires a
more massive disk.  The flattening is also degenerate with the halo
Einstein radius: massive halos with high $\RhE$ need to be more
flattened to reproduce the overall ellipticity of the projected mass
as constrained by the lensing data.  The scale radius of the
dark-matter halo is not constrained, as expected since lensing only
probes the distribution in the radial range spanned by the images,
i.e., $\sim$$0\farcs5$, or $\sim$$4\kpc$.  Nonetheless, a small value of
$\lesssim 1''$ is rejected by the data at 95\% CI.

\begin{figure}[t]
\centering
\includegraphics[width=0.47\textwidth]{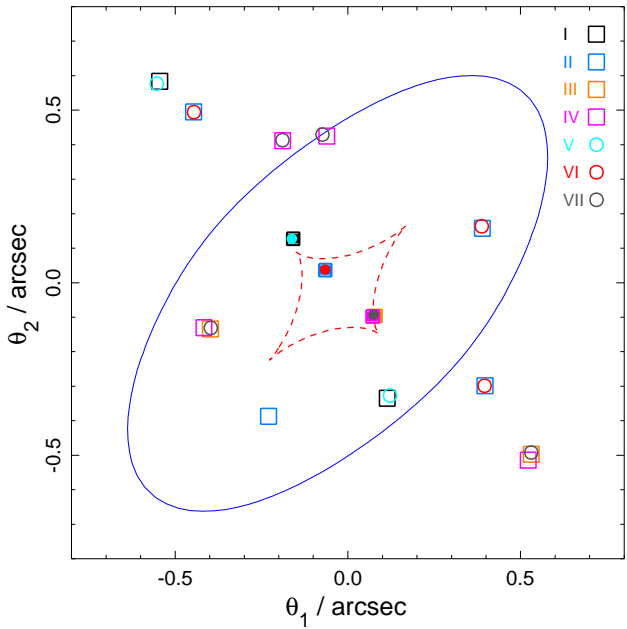}
\caption{\label{fig:critcaus} The critical curves (solid) and
  caustics (dashed) of the most probable model based on lensing data
  only. The open symbols are the observed image positions that are
  labeled according to Table \ref{tab:image-positions}:
  squares are the images positions in \citet{Cohn++01}, and circles
  are the global VLBI image positions.  The
  corresponding solid symbols are the modeled source positions. The 
  three groups of source positions, which have one group outside the
  astroid caustics (with image multiplicity of 2) and two groups
  inside the caustics (with image multiplicity of 4), form the
  10-image system.  The model reproduces the observed image positions
  to within $0\farcs02$, except for the merging components 2a and 2b (in
  systems IV and VII) which are reproduced to within $0\farcs06$ due to
  the higher positional uncertainty from the high magnification.} 
\end{figure}

The most probable lensing model (with highest posterior PDF) has a
reduced $\chi^2=0.9$.  We show in Figure \ref{fig:critcaus}
the critical curves (solid) and caustics (dashed) of the most probable
lensing model.  The open symbols are the observed image positions, and
the solid symbols are the modeled source positions.  The figure
illustrates the configuration of the 10-image system in relation to
its 3-component source (the source positions are clustered into three
groups).  The first group of sources is outside the astroid caustics
and produces components 1a and 8. The second group is inside the
astroid caustics and produces components 1, 3, 4 and 6.  The third
group is near a fold of the caustics and produces components 2, 5 and
7.

\section{Combining Kinematics and Lensing}
\label{sec:L+D}
\begin{figure}[t]
  \centering
  \includegraphics[width=0.47\textwidth, clip]{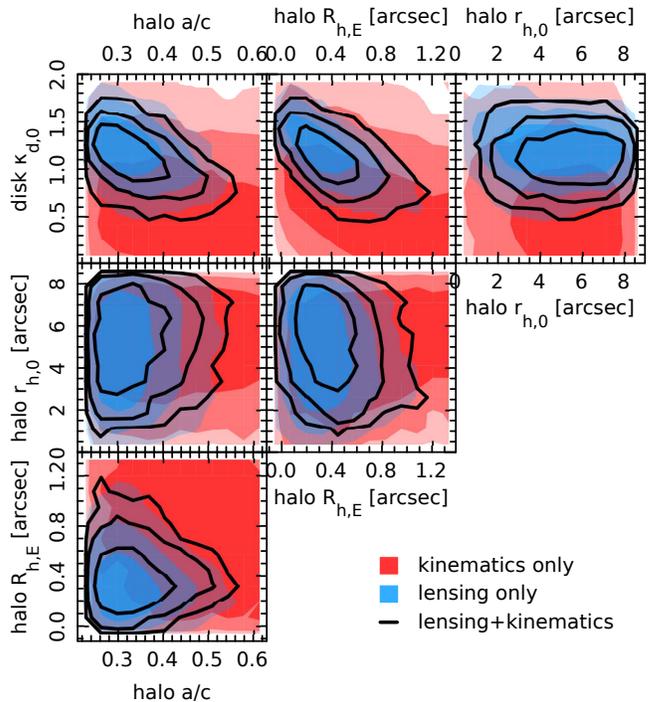}
  \caption{ \label{fig:PDF_LensDyn} Marginalized posterior PDF for the
    halo flattening a/c, halo Einstein radius $R_{\rm h,E}$, halo scale
    radius $r_{\rm h,0}$, and disk strength $\kappa_{\rm d,0}$ based on
    kinematics only (red), lensing only (blue), and joint lensing and
    kinematics (black).  The three shaded areas show the 68.3\%, 95.4\%
    and 99.7\% credible regions. }
\end{figure}

In this section, we present results on the mass distribution for the
spiral galaxy based on the kinematics and lensing data sets.  
Since the two data sets are independent, the likelihood is 
\be
P(\data_{\rm D}, \data_{\rm L} |\paramvec) = P(\data_{\rm D}
|\paramvec)\, P(\data_{\rm L} |\paramvec),
\ee
where the kinematics and lensing likelihoods on the right-hand side are
given by equations (\ref{eq:dynlike}) and (\ref{eq:lenslike}),
respectively. 
Figure~\ref{fig:PDF_LensDyn} shows the result of {\sc MultiNest} sampling of
the posterior.  The reduced $\chi^2$ of the joint data set is $0.8$.
The marginalized parameters with 68\% CI are listed in Table 
\ref{tab:par_LensDyn}.

\begin{table*}
\caption{Marginalized parameters (68\% CI)}
\label{tab:par_LensDyn}
\begin{center}
\begin{tabular}[b]{c c c c c c c c c c}

\hline
disk $q_{\rm d}$ & disk $\kappa_{\rm d,0}$ & bulge $M_{\rm b}$ & halo
$a/c$ & halo $q_{\rm h}$ & halo $\phi_{\rm h}$ & halo $\RhE$ &
halo $r_{\rm h,0}$& $\gamma_{\rm ext}$ & $\phi_{\rm ext}$ \\ 
 &  & ($10^{10}\,{\rm M}_{\sun}$) &  &  & ($\degr$)  & (arcsec) &
 (arcsec) & & ($\degr$) \\
\hline
$0.53\pm0.03$ & $1.1\pm0.2$ & $1.6\pm0.3$ & $0.33^{+0.07}_{-0.05}$ &
$0.67\pm0.04$ & $137\pm2$ & $0.4\pm0.2$ & $5\pm2$ & $0.02\pm0.01$ &
$103\pm14$ \\ 
\hline
\end{tabular}
\end{center}
Notes. The marginalized model parameters for the 3-component mass
distribution constrained by lensing and kinematics.  Columns 1--2 are
the projected disk axis ratio and strength.  Column 3 is the bulge
mass.  Columns 4--8 are the halo flattening, projected axis ratio,
projected position angle, Einstein radius and scale radius.  Columns
9--10 are the external shear strength and position angle.
\end{table*}

\subsection{Breaking the disk-halo degeneracy}
\label{sec:L+D:disk-halo-degen}
Although the kinematics constraints are significantly weaker than the
lensing constraints, the $\kappa_{\rm d,0}$--$R_{\rm h,E}$ panel (top-middle)
in Figure \ref{fig:PDF_LensDyn} shows that the kinematics and lensing
contours are tilted at different angles.  Thus, the combination of the
two in principle breaks the disk-halo degeneracy.  In the case of
\ourlens, most of the constraints on the mass distribution come from
the lensing data: the 10 radio images, which both span a large range of
radii and have positional uncertainties of only a few mas, provide a 
strong leverage in
discerning contributions from the disk, bulge and halo.  This is in contrast
to typical lenses with only two or four images which result in stronger disk-halo
degeneracies \citep[e.g.,][]{Trott++10}.  On the other hand, the
SNR of the rotation curve data for \ourlens\ is only modest given the
high lens redshift of $z_{\rm l}=0.755$.  Nonetheless,
the kinematics data in \ourlens\ are informative in excluding the high
disk masses allowed by lensing, which has interesting consequences
that are discussed in Section \ref{sec:IMF}.

A disk is considered to be maximal if the fractional contribution to
the circular velocity of the disk at $2.2\,R_{\rm d}$ is $v_{\rm
  d}(2.2\,R_{\rm d})/v_{\rm tot}(2.2\,R_{\rm d})=0.85\pm0.1$
\citep{Sackett97}.  Based on our lensing and kinematics model, \ourlens\
has $v_{\rm d}(2.2\,R_{\rm d}) = 248^{+22}_{-26}\kms$ and $v_{\rm
  tot}(2.2\,R_{\rm d}) = 326\pm8\kms$.  In comparison to the Milky
Way's circular velocity at the position of the sun, $v_{\rm
  c}(R_0)=(219\pm20)\,R_0/(8\,{\rm kpc})\kms$ \citep{Reid++99}, the
spiral galaxy in \ourlens\ is significantly more massive.
The resulting disk contribution in \ourlens\ of
$v_{\rm d}(2.2\,R_{\rm d})/v_{\rm tot}(2.2\,R_{\rm
  d})=0.76^{+0.05}_{-0.06}$ suggests that the disk is marginally 
submaximal.

\subsection{Shape and profile of the dark matter halo}
\label{sec:L+D:haloshape}

In the first column of Figure \ref{fig:PDF_LensDyn}, we see that the
dark matter halo in \ourlens\ is oblate with
$a/c=0.33^{+0.07}_{-0.05}$ in order to fit 
to the lensing observations.  The axis ratio of the projected surface
mass density of the halo is $0.67\pm0.04$, which is slightly rounder
than the axis ratio of projected disk of $0.53\pm0.03$.  It appears
that only projected mass distributions of the lens with axis ratios of
$\sim$$0.6$ are consistent with the lensing data.  If we consider the
subset of the {\sc MultiNest} sample with $a/c\,\sim\,0.5$ (which corresponds to
a projected axis ratio of $\sim$$0.75$), then the lensing reduced
$\chi^2$ increases from $0.8$ to $2.4$ due to misfits in
systems II, V, VI and VII in Table \ref{tab:image-positions}.
The global VLBI data with better positional accuracies and, in
particular, the four-image systems II and VI enforce the high
ellipticity in the projected surface mass density and hence the highly
oblate halo.  The high ellipticity is robust against assumptions on
the dark matter halo profile.  In particular, when we replace the NFW
halo with either a singular or cored isothermal profile, the radio
data still require a high ellipticity for the halo. 

$N$-body simulations of dark matter halos indicate that the halos are
typically triaxial and the axis ratio between the short and long axis
(i.e., $a/c$ in our notation) is $\sim$$0.3$ to $0.9$ (95\%CL) for halo
masses of $10^{11}-10^{12}\,{\rm M}_{\sun}$ \citep[e.g.,][]{Bett++07,
  Maccio++08}.  Simulations with baryons find that the central galaxy
tends to make the triaxial halo essentially oblate and less
flattened \citep[e.g.,][]{Abadi++09}.  Our study of \ourlens\ with the
resulting $a/c\,\sim\,0.3$ for the oblate NFW halo suggests that baryons
are effective at making the halo oblate in the inner regions of the
galaxy.  

Since the kinematics and 
lensing data probe the inner $\sim$$1''=7.5\kpc$ of the galaxy, it is
not surprising that the scale
radius $r_{\rm h,0}$ of the dark matter halo in our model (where the density
transitions from $\rho\sim r^{-1}$ to $r^{-3}$) is not constrained but
has a lower limit of $2\farcs1 = 16\kpc$ (95\% CI).  The virial radius is
difficult to extrapolate from the model given the uncertain scale
radius.  For galaxies with total mass of $\sim$$10^{12}\,{\rm M}_{\sun}$
(applicable to \ourlens), typical virial radii are 
$\sim$$250\kpc$.  This implies a concentration of $\lesssim 16$ that is
consistent with concentration-mass relations for galaxy-scale halos
\citep[e.g.,][]{Maccio++08,Duffy++10}.  To better constrain
$r_{\rm h,0}$, a rotation curve that extends to larger radii is needed.

\subsection{Dark matter mass fraction}
\label{sec:L+D:fDM}

\begin{figure}[t]
  \centering
  \includegraphics[width=0.33\textwidth, angle=270]{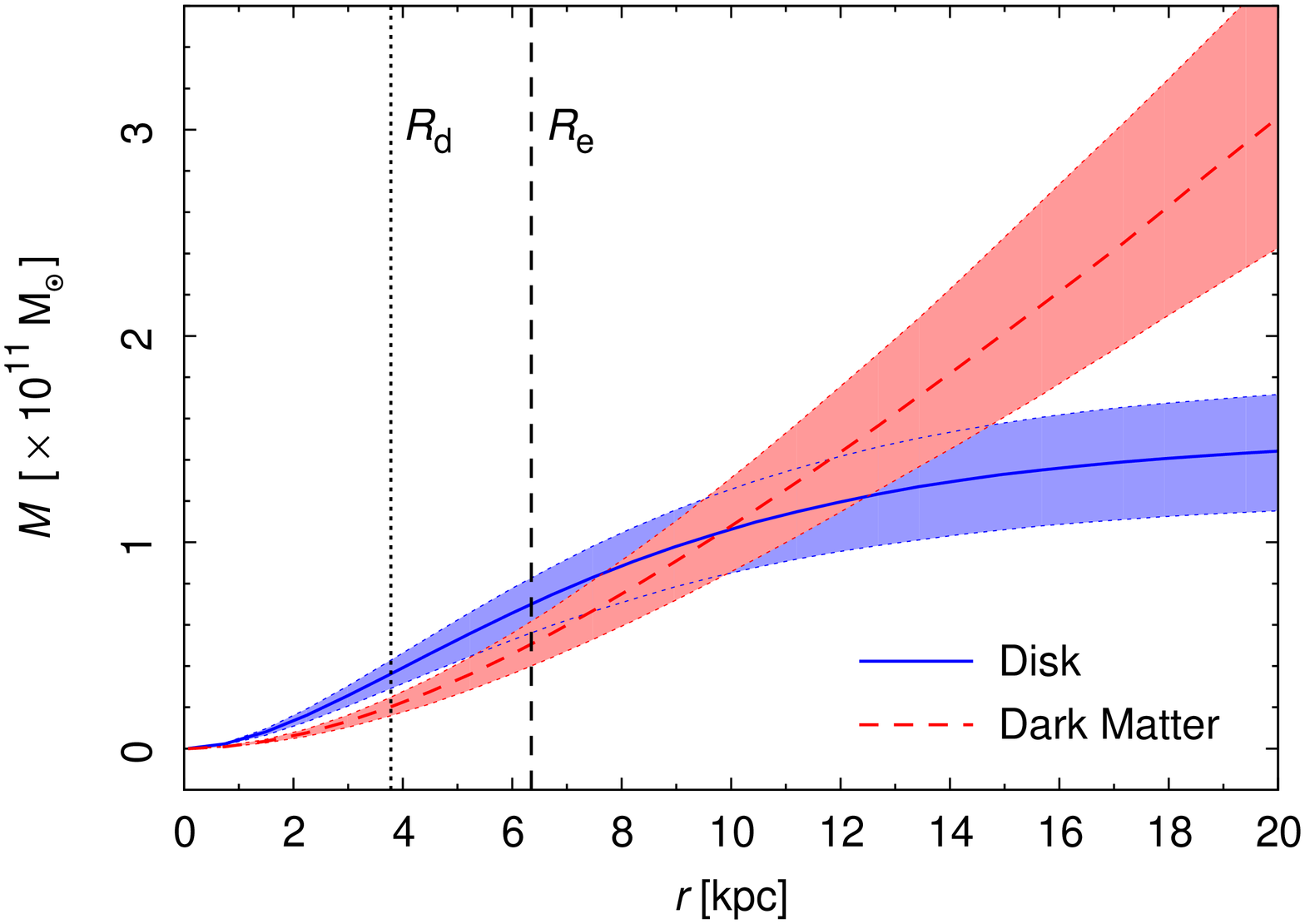}
  \includegraphics[width=0.33\textwidth, angle=270]{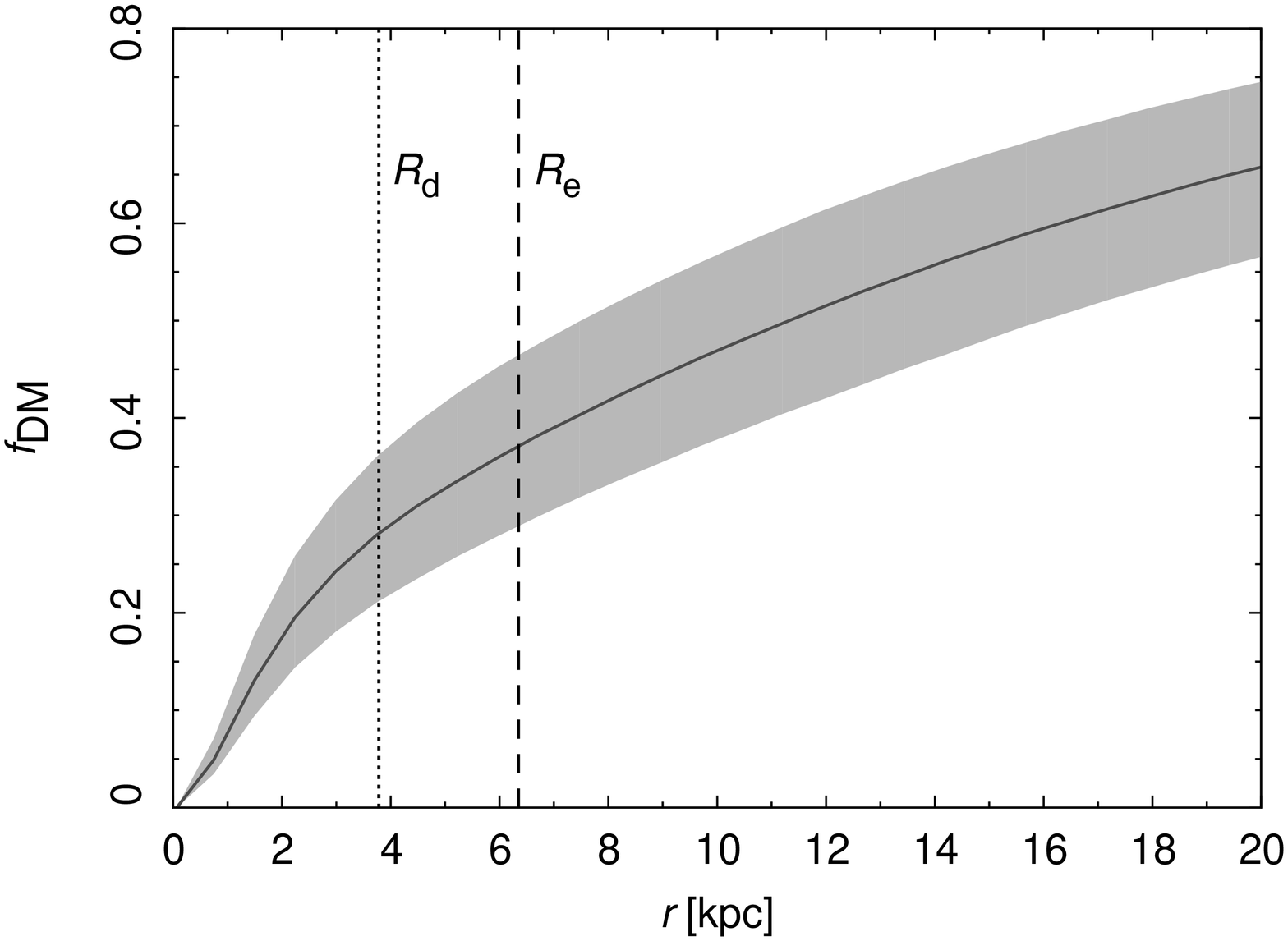}
  \caption{ \label{fig:M+fDMvsR} Masses and dark matter mass
    fractions from the lensing and kinematics modeling.  Top panel:
    spherical enclosed mass of the disk and halo components.  The
    bulge is modeled as a point mass of $\mbulge$.  Bottom
    panel: the dark matter mass fraction within a sphere.  The dotted
    (dashed) vertical line is the scale (effective) radius of the
    disk.}
\end{figure}

From our 3-component
model of the galaxy, we can determine the fraction of dark matter as a
function of radius by integrating the mass enclosed within spherical
radii.  Since we employ a parametrized model and the parameters of
the model are constrained by the lensing and kinematics data, the
enclosed mass can be computed for all radii.  At small and large radii
where there is no (or low SNR) data, the enclosed mass is effectively
extrapolated based on the form of the parametrized model and the
measured values of the mass parameters.
The top panel in Figure \ref{fig:M+fDMvsR} shows the mass
enclosed for the disk and the dark matter halo.  The bulge, which is
unresolved and modeled
as a point, has mass $M_{\rm b}=(1.6\pm0.3)\times10^{10}\,{\rm
  M}_{\sun}$ for all radii.  In the inner $10\kpc$
of the galaxy, we see that the disk dominates in mass, and beyond
that, the dark matter halo dominates.

The bottom panel shows the dark matter mass fraction within a sphere
of radius $r$.  The rise in the dark matter fraction is similar to the
analysis of the spiral lens SDSS\,J2141$-$0001 by \citet{Dutton++11}.
The dark matter mass fraction within 2.2 disk scale lengths for
\ourlens\ is $f_{\rm DM,2.2}=0.43^{+0.10}_{-0.09}$. 
Within the effective radius, the mass fraction is $f_{\rm
  DM,e}=0.37^{+0.09}_{-0.08}$, which is consistent 
with the ranges of values found in previous lensing and kinematics
analyses of early-type galaxies \citep[e.g.,][]{Auger++10,
  Barnabe++11}.

\subsection{Implications for the IMF}
\label{sec:IMF}

The shapes of the low- and high-mass end of the stellar IMF are
difficult to determine observationally since low mass stars are
intrinsically faint and high-mass stars are low in abundance.
In this section, we describe the estimation
of the stellar mass in \ourlens\ based on SPS
models for \citet{Chabrier03} and \citet{Salpeter55}
IMFs.  We compare these masses to the independently measured disk
mass from lensing and kinematics to learn about the IMF.

\begin{figure}[t]
\centering
\includegraphics[width=0.48\textwidth]{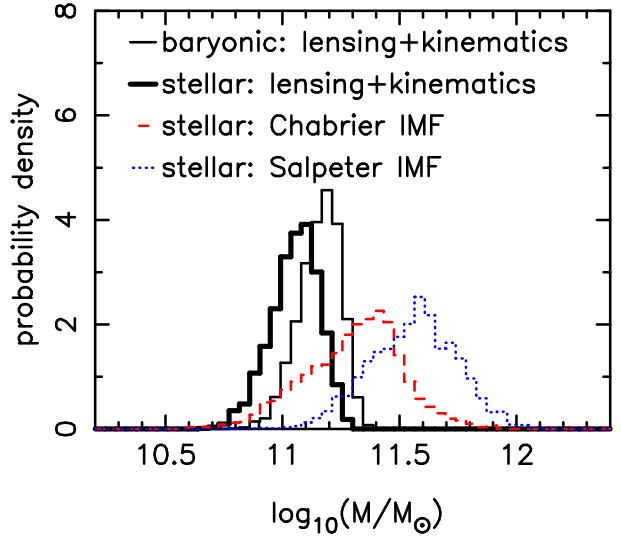}
\caption{ \label{fig:Mdisk} Comparison of the disk mass from the 
  lensing and kinematics analysis, and the stellar mass from
  photometry and stellar population synthesis with Chabrier or
  Salpeter IMFs.  A Chabrier-like IMF is preferred to a
  Salpeter-like IMF by a factor of $\imfevid$ when a $20\%\pm10\%$
  contribution in mass from the cold gas is assumed.}
\end{figure}

Using the broadband photometries in Table \ref{tab:lensmag}, we follow
\citet{Auger++09} to infer the stellar mass for the disk based on
composite stellar population synthesis models from
  \citet{BruzualCharlot03}.  These models have been employed in many
  studies including SDSS galaxies \citep[e.g.,][]{Kauffmann++03,
    Gallazzi++05} and spiral lens analyses similar to ours
  \citep[e.g.,][]{Dutton++11}. Dust extinction (intrinsic) is taken
  into account in these models, and our multiband photometries allow a
  good handle on dust with only a small broadening of
  uncertainties \citep[e.g.,][]{BelldeJong01}.  Our inference of
  the stellar mass provides
uncertainties that incorporate parameter degeneracies in the stellar
population models (e.g., between age and metallicity).
In Figure \ref{fig:Mdisk}, we plot the 
inferred stellar mass in red-dashed (blue-dotted) assuming Chabrier
(Salpeter\footnote{We use $0.1\,{\rm M}_{\sun}$ as the lower mass limit
  for the Salpeter IMF.}) IMF.  These two IMFs cover the range
applicable to spiral galaxies \citep{BelldeJong01}.  

The lensing and kinematics analysis provides an independent measurement
of the total disk mass of $\log_{10}(M_{\rm disk}/{\rm M}_{\sun}) =
11.17^{+0.08}_{-0.10}$ that includes stars and gas (thin solid 
curve in Figure \ref{fig:Mdisk}).  To extract the stellar mass
contribution, we assume that the cold gas accounts for $20\pm10\%$ of
the total mass \citep[e.g.,][]{DuttonVandenBosch09} and is distributed
like the stars.  This provides an upper limit in the mass contribution
of the gas to the disk (in comparison to scenarios where the gas is
more extended than the stars).  For each sample in the posterior PDF
of the disk mass, we draw a random number $f$ from a Gaussian
distribution centered on $0.2$ with a standard deviation of $0.1$ that
is truncated between [0,1], and subtract the fraction $f$ from the
total disk mass.  The gas-subtracted disk mass from the lensing and
kinematics model is $\log_{10}(M_{*}/{\rm M}_{\sun}) =
11.06^{+0.09}_{-0.11}$, and the distribution is shown by the thick solid
curve in Figure \ref{fig:Mdisk}. 

Comparing the thick solid curve to the dashed and dotted curves in the
figure, we see that our mass model of \ourlens\ favors a Chabrier-like
IMF.  We can quantitatively rank the two IMFs by computing the
Bayesian evidence which is the integral under the product of the PDF
from lensing and kinematics (thick solid) and the PDF from the SPS (dashed or
dotted).  The ratio of the Bayesian evidence for Chabrier to Salpeter
is 7.2; in other words, the probability of the IMF being Chabrier is
7.2 times higher than the probability of the IMF being Salpeter for
\ourlens.

Using the Chabrier IMF, we obtain the rest-frame stellar mass-to-light
ratio of the disk to be $M_{*}/L_{\rm V}=(0.6\pm0.3) {\rm M_{\sun}/L_{\rm
    V,\sun}}$.  By passively evolving to $z=0$, we obtain the present
day stellar mass-to-light ratio of $M_{*}/L_{\rm V}=(1.7^{+1.3}_{-0.9})
{\rm M_{\sun}/L_{\rm V,\sun}}$, in agreement with typical values found
for local galaxies \citep[e.g.,][and references therein]{Trott++10,
 vandeVen++10, CourteauRix99, vanderKruitFreeman11}.

Our finding of a preference toward a Chabrier-like IMF for the spiral
galaxy is in agreement with the results in \citet{Dutton++11} and
\citet{Ferreras++10}. 
Nonetheless, studies of massive elliptical galaxies favor
Salpeter-like IMFs
\citep[e.g.,][]{Treu++10,Auger++10b,vanDokkumConroy10, Spiniello++11}.
This supports  
a non-universal IMF for galaxies that is dependent on mass and/or
Hubble type.

\section{Conclusions}
\label{sec:conclude}
We disentangled the distributions of baryons and dark matter in the
spiral galaxy \ourlens\ by using the newly acquired global VLBI
observations, AO-assisted NIRC2 imaging, rotation curve, and source
redshift.  We constructed an axisymmetric 3-component mass model for
the galaxy with an exponential disk, a point mass to approximate the
unresolved bulge, and a NFW dark matter halo.  Parameters of this
model were constrained by a combined lensing and kinematics analysis.
Based on this study, we conclude the following 
\begin{itemize}
\item The image positions of the radio source span a range
  of radii and provide strong constraints on the 3-component mass 
  distribution of the lens.
\item The fractional contribution of the disk to the total circular
  velocity at $2.2\,R_{\rm d}$ is $0.76^{+0.05}_{-0.06}$, suggesting
  that the disk is marginally submaximal.
\item The oblate dark matter halo needs to be highly flattened with
  $a/c\,\sim\,0.3$ in order to fit to the radio observations.  
\item The lensing and kinematics data sets probe the inner $\sim$$10\kpc$
  region of the mass distribution and place a lower limit of $16\kpc$
  (95\% CI) for the scale radius of the NFW halo.
\item The dark matter mass fraction inside a sphere increases as a
  function of radius.  The mass fraction within $2.2\,R_{\rm d}$ is
  $f_{\rm DM,2.2}=0.43^{+0.10}_{-0.09}$, and within the effective 
  radius is $f_{\rm DM,e}=0.37^{+0.09}_{-0.08}$.
\item The total stellar mass of the disk based on the lensing and
  kinematics data sets is $\log_{10}(M_{*}/{\rm M}_{\sun}) =
  11.06^{+0.09}_{-0.11}$, assuming that the cold gas is distributed
  like the stars.  
\item Based on the lensing and kinematics measurement of the disk mass,
  the Chabrier IMF is preferred to the Salpeter IMF by a probability
  factor of 7.2.

\end{itemize}

The sample of spiral lenses has been growing rapidly in recent years
thanks to dedicated surveys \citep[e.g.,][]{Marshall++08, Sygnet++10,
  Treu++11}.  While most of the lenses do not have sources that are
radio loud, spatially extended lensed images can also be used to
constrain the mass distribution \citep[e.g.,][]{Dutton++11}.  The
combined lensing and kinematics modeling methods we have developed are
general, and important insights into the interactions between baryons
and dark matter in the formation and evolution of spiral galaxies can
be derived from a sample of lenses using these techniques.

\begin{acknowledgements}
We would like to thank Matteo Barnab\`e, Aaron Dutton, Phil
Marshall for useful discussions, and the anonymous referee for
the constructive comments that improved the presentation of the paper.
SHS and TT acknowledge support from the Packard Foundation through a
Packard Research Fellowship to TT. SHS is supported in part through
HST grants 11588 and 10876.  CDF acknowledges support from NSF-AST-0909119.
LVEK is supported in part by an NWO-VIDI program subsidy (project
number 639.042.505). 
This work was supported in part by the Deutsche
Forschungsgemeinschaft under the Transregio TR-33 ``The Dark
Universe''. 
The National Radio Astronomy Observatory is a facility of the National
Science Foundation operated under cooperative agreement by Associated
Universities, Inc.
The European VLBI Network is a joint facility of European, Chinese,
South African and other radio astronomy institutes funded by their
national research councils.  
Some of the data presented in this paper were obtained at the
W.M. Keck Observatory, which is operated as a scientific partnership
among the California Institute of Technology, the University of
California and the National Aeronautics and Space Administration. The
Observatory was made possible by the generous financial support of the
W.M. Keck Foundation.  The authors wish to recognize and acknowledge
the very significant cultural role and reverence that the summit of
Mauna Kea has always had within the indigenous Hawaiian community.  We
are most fortunate to have the opportunity to conduct observations
from this mountain.  

\end{acknowledgements}

\bibliographystyle{apj}
\bibliography{ms}

\end{document}